\documentclass[a4paper,fleqn,usenatbib,useAMS]{mnras}

\usepackage{graphicx}   
\usepackage{amsmath}    
\usepackage{amssymb}    
\usepackage{multicol}        
\usepackage{bm}         
\usepackage{pdflscape}  
\usepackage{comment}
\usepackage{lineno}

\usepackage[T1]{fontenc}
\usepackage{ae,aecompl}

\title{Dissecting the morphology of star forming complex S193}
\author[Pandey et al.]
{Rakesh Pandey$^{1}$\thanks{pandey.rakesh405@gmail.com}, Saurabh Sharma$^{2}$,Lokesh Dewangan$^{1}$,D. K. Ojha$^{3}$, Neelam Panwar$^{2}$ \newauthor
Arpan Ghosh$^{2}$, Tirthendu Sinha$^{4}$, Aayushi Verma$^{2}$, and Harmeen Kaur$^{5}$ \\
$^{1}$Physical Research Laboratory, Navrangpura, Ahmedabad - 380 009, India.\\
$^{2}$Aryabhatta Research Institute of Observational Sciences (ARIES), Manora Peak, Nainital, 263 002, India, saurabh@aries.res.in\\
$^{3}$Tata Institute of Fundamental Research, Homi Bhabha Road, Colaba, Mumbai - 400 005, India\\
$^{4}$Satyendra Nath Bose National Centre for Basic Sciences (SNBNCBS), Block-JD, Sector-III, Salt Lake, Kolkata-700 106, India\\
$^{5}$Center of Advanced Study, Department of Physics DSB Campus, Kumaun University Nainital, 263002, India\\
}

\date{Accepted XXX. Received YYY; in original form ZZZ}

\pubyear{2023}

\begin{document}
\label{firstpage}
\pagerange{\pageref{firstpage}--\pageref{lastpage}}
\maketitle

\begin{abstract}
{
We have studied a star-forming complex S193 using near-infrared (NIR) observations and other archival data covering optical to radio wavelengths. We identified stellar clusters in the complex using the NIR photometric data and estimated the membership and distance of the clusters. Using the mid-infrared (MIR) and far-infrared (FIR) images, the distribution of the dust emission around H\,{\sc ii} regions is traced in the complex. The $Herschel$ column density and temperature maps analysis reveal 16 cold dust clumps in the complex. The H$\alpha$ image and 1.4 GHz radio continuum emission map are employed to study the ionised gas distribution and infer the spectral type and the dynamical age of each H\,{\sc ii} region/ionised clump in the complex.
The $^{12}$CO(J =3$-$2) and $^{13}$CO(J =1$-$0) molecular line data hint at the presence of two velocity components around [-43,-46] and [-47,-50] km/s, and their spatial distribution reveals two overlapping zones toward the complex. By investigating the immediate surroundings of the central cluster [BDS2003]57 and the pressure calculations, we suggest that the feedback from the massive stars seems responsible for the observed velocity gradient and might have triggered the formation of the central cluster [BDS2003]57.}
\end{abstract}


\section{Introduction} \label{sec1}

Understanding the formation process of massive stars has been one of the leading research goals of many current and past studies \citep{2016A&A...592A..54A,2017ApJ...835..142T,2018NatAs...2..478M,2018ApJ...859..166F,2021PASJ...73S.405F}  as they have a profound effect on the Galactic evolution. The formation processes of massive stars are also less understood as they are of a short time scale and are less frequent than the lower mass stars.  
Since massive stars start affecting their natal environment
soon after their formation, these processes are even more complex \citep{2011AJ....141..123A}. Stellar collision  \citep{1998MNRAS.298...93B}, competitive accretion \citep{2001MNRAS.323..785B} and monolithic collapse of a dense cloud are considered in the literature as plausible scenarios to explain the formation process of massive stars. 
In recent years, cloud-cloud collision (CCC) has emerged as a very promising mechanism to explain massive star formation, as demonstrated by numerous theoretical and observation works \citep{2009ApJ...696L.115F,2017ApJ...851..140D,2018ApJ...859..166F,2021PASJ...73S.405F,2021ApJ...913...14L}. 
The CCC model was first conceptualized by \citet{1992PASJ...44..203H} where they realized 
the possibility of massive star formation in the compressed layer between one small and another large cloud formed via supersonic collision. 
Recently, \citet{2021PASJ...73S.405F}, with a detailed magneto-hydrodynamic simulation, show that the dense gas cores form in the compressed layer between two clouds, which preferentially triggers the formation of O and early B-type stars. 
The powerful winds, jets, and outflows from the massive stars change the physical conditions such as temperature, density, and turbulence 
in the surrounding molecular cloud \citep{2009ApJ...694L..26G}. For example, H\,{\sc ii} regions are formed through the expansion of 
ionised gas in the surrounding cloud due to the highly energetic UV radiation and winds from these massive stars. 
As most of the star formation occurs in a group or cluster,
massive stars are often associated with young star clusters and H\,{\sc ii} regions in a star-forming region. 
The feedback from the massive stars thus plays a vital role in governing the star formation process around them. 
It can push forward and increase the star formation rate, known as positive feedback or slow down or terminate star formation with negative feedback. The outcome depends not only on the process but the surrounding environment itself \citep{2017MNRAS.467..512S}.

Lying between the star-forming regions W4 and W5, the S193 complex ($\alpha$$_{2000}$ =02$^{h}$47$^{m}$25$^{s}$, $\delta$$_{2000}$ = 61$\degr$56$\arcmin$57$\arcsec$) 
consists of three  H\,{\sc ii} regions Sh 2-192 (hereafter, S192), Sh 2-193 (hereafter, S193) and Sh 2-194 (hereafter, S194) \citep{2008A&A...484..361Q,1982ApJS...49..183B}.
This region also hosts three previously identified massive stars, i.e., S192-1 (B2.5V), S193-1 (B2.5V) \& S193-3 (B1.5V) \citep{2007A&A...470..161R}, along with
three IRAS sources, IRAS 02435+6144 ($\alpha$$_{2000}$ =02$^{h}$47$^{m}$25$^{s}$, $\delta$$_{2000}$ = 61$\degr$56$\arcmin$57$\arcsec$),
IRAS 02437+6145 ($\alpha$$_{2000}$ =02$^{h}$47$^{m}$40.4$^{
s}$, $\delta$$_{2000}$ = 61$\degr$45$\arcmin$55$\arcsec$), IRAS 02439+6143 ($\alpha$$_{2000}$ =02$^{h}$47$^{m}$53.6$^{s}$, $\delta$$_{2000}$ = 61$\degr$56$\arcmin$01$\arcsec$) and a Be star D75b ($\alpha$$_{2000
}$ =02$^{h}$47$^{m}$35$^{s}$, $\delta$$_{2000}$ = 61$\degr$55$\arcmin$30$\arcsec$) \citep{2016A&A...591A.140G}.
Two  open clusters, i.e, [BDS2003]57 \citep{2007A&A...470..161R} and Teutsch
162 (hereafter, T162) \citep{2008A&A...484..361Q}, are also part of this region.
In the S193 complex, \citet{2011AJ....141..123A} studied the molecular gas in a velocity range of [-44,-50] km/s (see also \citet{2008A&A...484..361Q}) and identified ten molecular clumps. The velocities of the ionised gas toward S192, S193, and S194 were reported to be -44.7, -49.6, and -44.7 km/s, respectively \citep{2014ApJS..212....1A}.
The S193 complex is thus an exciting region to investigate the star formation processes involving massive stars, young star clusters, and molecular clumps.
In this paper, we have done a comprehensive multiwavelength study of the S193 complex, which is presented as follows.
Section \ref{sec2} has information about the multiwavelength data sets used in this study and explains the data reduction procedures regarding our near-infrared (NIR) observations.
In Section \ref{sec3}, we analyzed the multiwavelength data sets to estimate the various morphological parameters and probed the physical environment 
around the S193 complex. In section \ref{sec4}, we have discussed the star formation scenario in the S193 complex and summarized our study in Section \ref{sec5}.

\section{Observation and data reduction} \label{sec2}
\subsection{Near-infrared (NIR) Imaging data} \label{nird}
We observed the S193 complex in broad-band NIR ($JHK$) filters using the TIFR Near Infrared Spectrometer and Imager (TIRSPEC)\footnote{http://www.tifr.res.in/$\sim$data/tirspec/}. 
The instrument is mounted on the 2~m Himalayan $Chandra$ Telescope (HCT), Hanle, Ladakh, India. \citet{2014JAI.....350006N} have provided the details of this instrument 
and its detector array specifications. TIRSPEC covers  $307^{\prime\prime}\times 307^{\prime\prime}$ of the sky in the imaging mode; thus it took four pointings
to cover the entire S193 complex ($\sim10^{\prime}\times 10^{\prime}$). 
In each pointing, we took five dithered positions with seven frames in each dithered position. The exposure time of frames in all three bands ($JHK$) is 20 secs. We have also observed the central embedded cluster `[BDS2003]57' (FOV $\sim$  $1^{\prime}\times 1^{\prime}$ ) in $JHK$ bands using the 
TIFR-ARIES Near Infrared Spectrometer \citep[TANSPEC;][]{2022PASP..134h5002S} mounted on the 3.6m Devasthal Optical Telescope (DOT), 
Nainital, India. We observed the source in six sets, each containing seven dithered positions with five frames in each position. The exposure time of frames in the K band is 20 secs, while for the J and H bands, it is 50 secs. The NIR data reduction, which includes cleaning the raw images, and performing photometry and astrometry, is done using the procedure explained in \citet{2020ApJ...891...81P}. Using the transformation equations below, we transformed our instrumental $JHK$ magnitudes into a standard Vega system.

\begin{equation}
(J-K)= (0.92\pm0.12)\times (j-k)+ (0.64\pm0.06)          
\end{equation}
\begin{equation}
(H-K)=(0.99\pm0.03)\times(h-k) + (-0.04\pm0.02)
\end{equation}
\begin{equation}
(K-k)= (-0.04\pm0.10)\times(H-K) +  (-4.85\pm0.02)
\end{equation}

In the above equations, $jhk$ and $JHK$ denote the instrumental and standard magnitudes, respectively. 
The coefficients in the above equations were generated using the stars common in our observations and the 2MASS catalogue
for one of the pointings in the TIRSPEC observations. Similarly, the coefficients were estimated separately for each pointing and
instrument. We made a combined catalogue that includes the TIRSPEC
and TANSPEC observations by considering only those stars with less than 0.1 mag photometric uncertainty. The magnitudes of the stars which are saturated in our observations are taken from the 2MASS catalogue. Table \ref{stats} provides the statistics of our NIR observations.  

\subsection{Archival data} \label{archiv}

The multiwavelength archival data sets used in the present study are tabulated in Table \ref{archilog}. The optical photometric data in the grizy bands (i.e., g, r, i, z, and y) were taken from the Pan-STARRS1 Surveys (PS1). The 5$\sigma$ limiting magnitudes of point sources in g, r,  i,  z, and y bands are 23.3,  23.2, 23.1, 22.3, and 21.4 mag, respectively \citep{2016arXiv161205560C}. We particularly used g and i band photometric data from the survey, to plot the colour-magnitude diagram (CMD, cf. Section \ref{dist}). We accesed the point-source catalogue of 2MASS survey in NIR $J$, $H$, and $K$ bands. The catalogue is 99 \% complete upto the $J$= 15.8 mag, $H$=15.1 mag, and $K$=14.3 mag \citep{2006AJ....131.1163S}, we used the 2MASS NIR data to generate the surface density map of the stars in the targeted region (cf. Section \ref{group}). The Gaia Early Data Release 3 (DR3) \citep{2021A&A...649A...1G} data are used in the present work for estimating the member stars of the clusters using their proper motion. Gaia DR3 provides astrometry, photometric, and radial velocity measurements of the nearly $\sim$ 1.8 billion sources, the data shows completeness of 99 \% upto the G $\sim$ 20 (for b $\sim$ 30$\degr$), and G $\sim$ 22 (for higher lattitudes).\\
We used Mid-infrared (MIR) and Far-infrared (FIR) images from $Spitzer$ and $Herschel$, respectively to trace the heated dust emission in the S193 complex, while Submillimetre (Sub-mm) images from $Herschel$ are effective in tracing the cold dust emission. The radio continuum image from the NRAO VLA Sky Survey (NVSS) along with the H$\alpha$ image from the INT Photometric H$\alpha$ Survey of the Northern Galactic Plane (IPHAS), are used in this work, to trace the distribution of the ionised gas (cf. Section \ref{ioni}). \textbf{We also used the $^{12}$CO(J =3$-$2) line data from the 15m JCMT telescope (James Clerk Maxwell Telescope) archive and the $^{13}$CO(J =1$-$0) data from the 13.7m Purple Mountain Observatory (PMO, China) acquired as a part of the Milky Way Imaging Scroll Painting (MWISP)project \footnote{http://english.dlh.pmo.cas.cn/ic/in/}. The molecular line data are used for tracing the kinematics of the molecular gas in the targeted region (cf. Section \ref{jcmK}).}
 
\begin{figure*}
\centering
\includegraphics[width=1\textwidth]{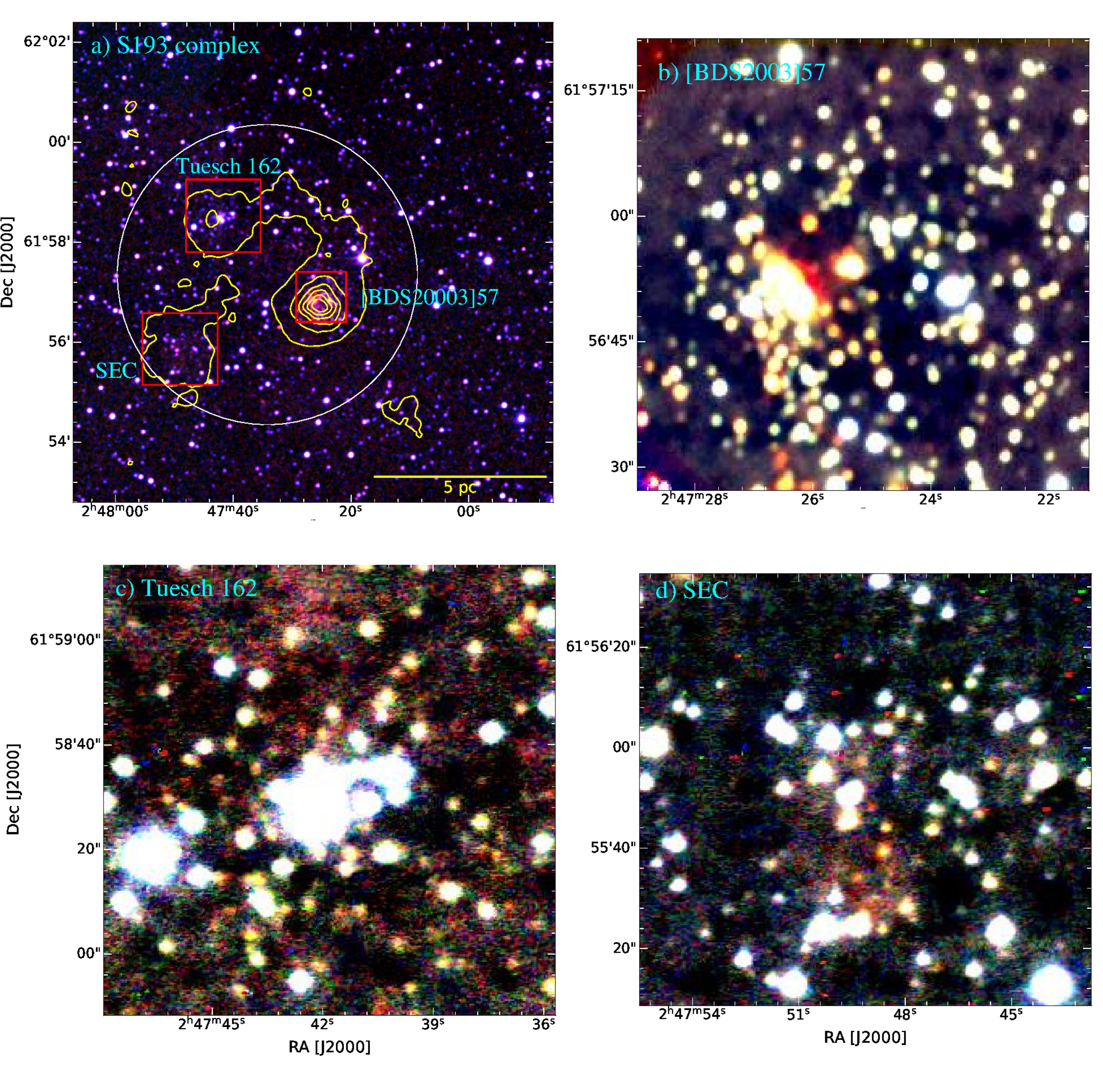}
\caption{\label{nirrgb} (a) Colour composite image of the S193 complex made using 2MASS $K$ (red), $H$ (green), $J$ (blue) band images. The surface density contours
from the present analysis (see Section \ref{group}) are overplotted with the yellow colour, while the core region of the three clusterings ([BDS2003]57, T162, and SEC) is marked with the red square. (b) TANSPEC $K$ (red), $H$ (green), and $J$ (blue) band colour composite image of the core region of the [BDS2003]57 clustering [FOV $\sim$  $56^{\prime\prime}\times 56^{\prime\prime}$]. TIRSPEC $K$ (red), $H$ (green) and $J$ (blue) band colour composite image of the core region of the T162 [FOV $\sim$  $86^{\prime\prime}\times 86^{\prime\prime}$] and the SEC [FOV $\sim$ $90^{\prime\prime}\times 90^{\prime\prime}$] are shown in (c) and (d), respectively.   
}
\end{figure*}

\begin{figure}
\centering
\includegraphics[width=8cm,height=4cm]{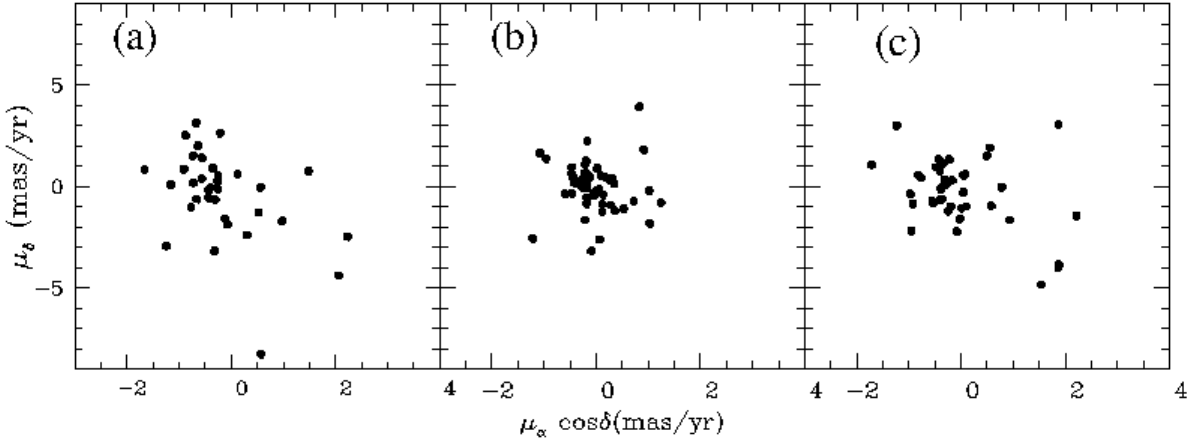}
\caption{\label{vpd} PM vector-point diagrams for [BDS2007]57 (a), Tuesch 162 (b), and SEC (c) clustering.
}
\end{figure}

\begin{figure}
\centering
\includegraphics[width=8cm,height=9cm]{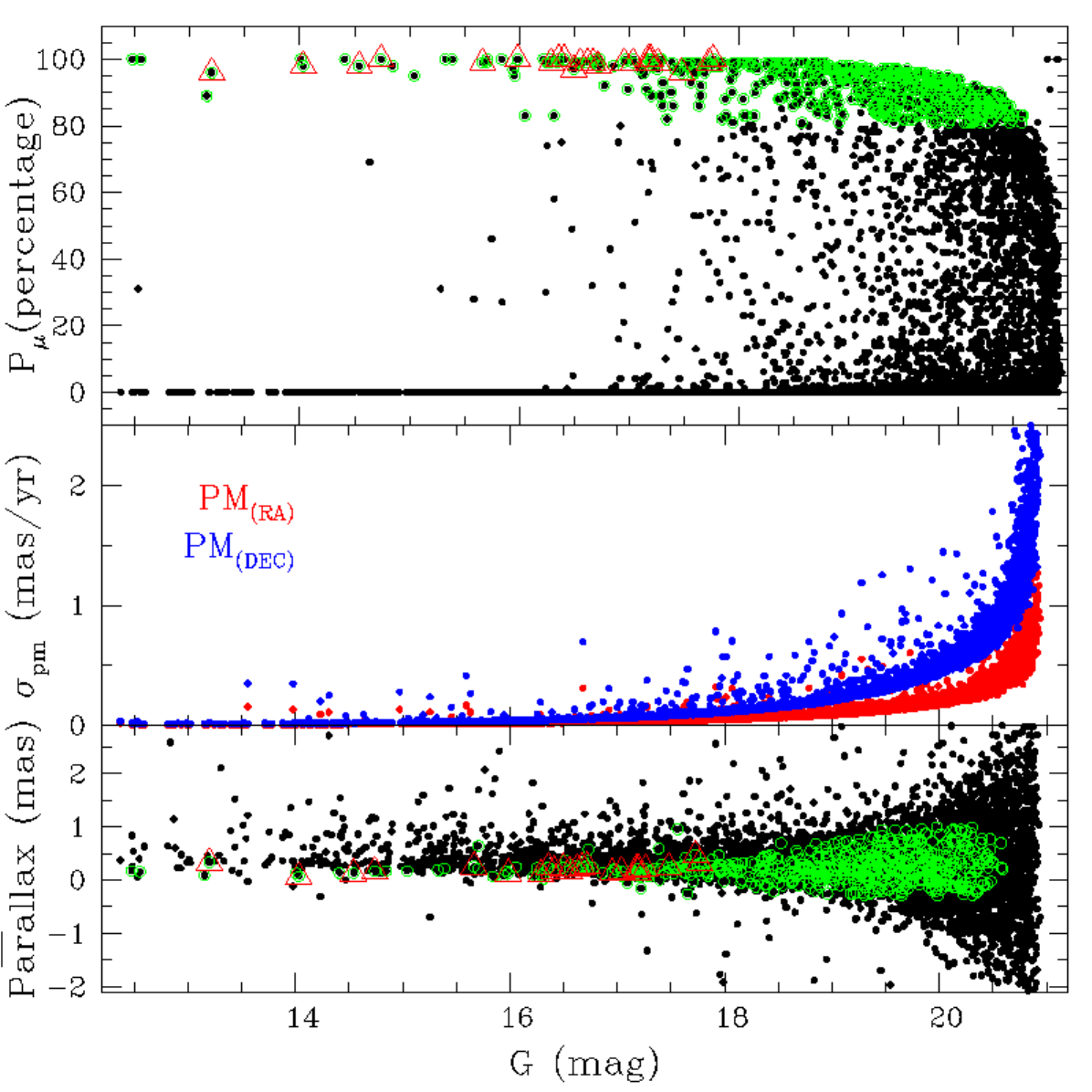}
\caption{\label{pm} Membership probability (P$_\mu$), PM errors ($\sigma_{PM}$) and parallax of all the stars within the clustering (circle having radius 3$^{\prime}$) identified in the S193 complex as a function of $G$ magnitude. The probable member stars (P$_\mu>$80 \%) are shown by green circles, while 24 members of the S193 complex having parallax values with good accuracy ({err $<$ 0.1} mas) are shown by red triangles.
}
\end{figure}

\begin{figure}
\centering
\includegraphics[width=0.45\textwidth]{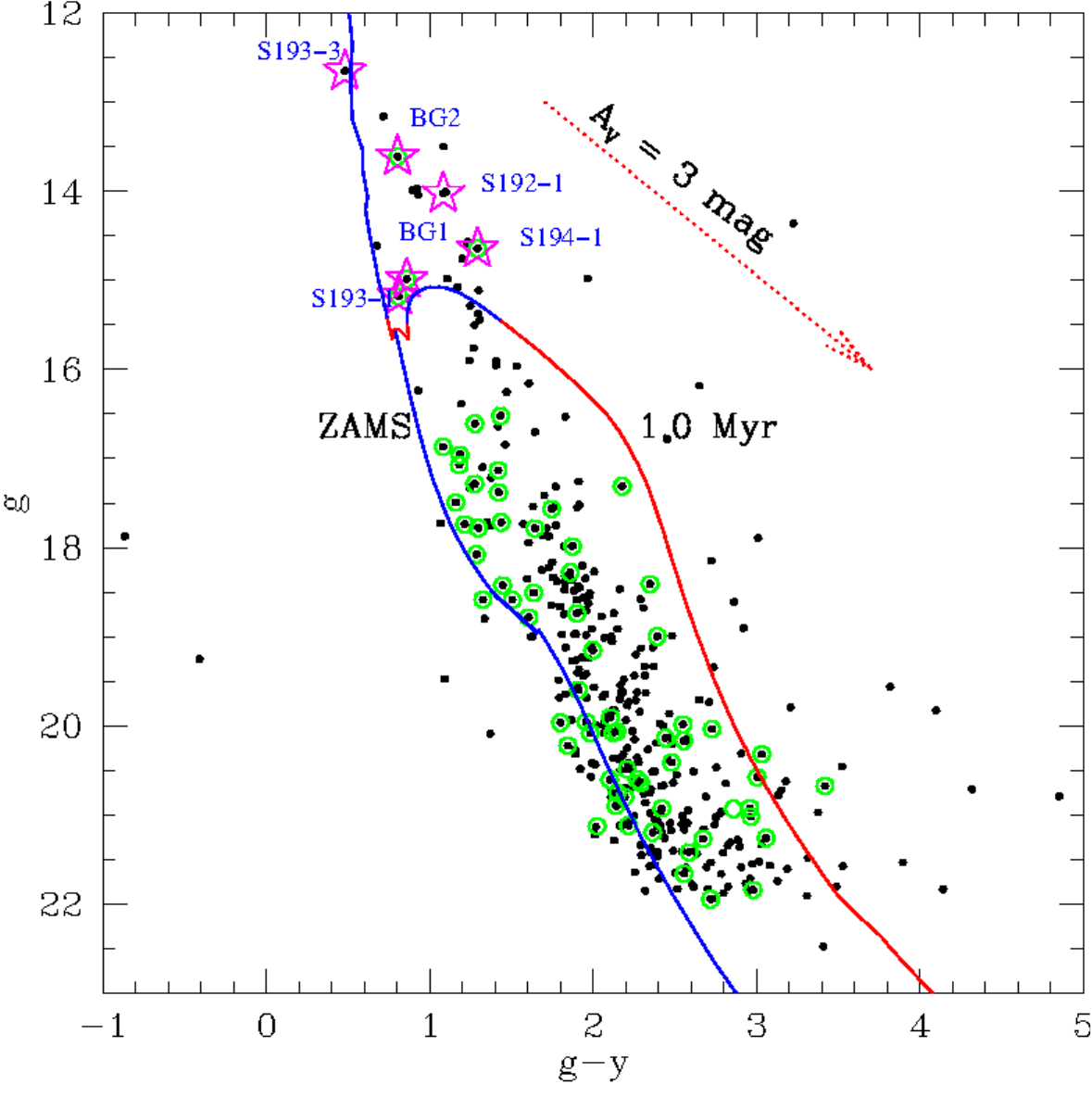}
\caption{\label{ccd} g vs (g-y) CMD, plotted using the PS1 data, the black dots represent the star within the cluster while green dots represent the member star identified in the present analysis (see section \ref{member}). The ZAMS and 1 Myr isochrone by \citet{2019MNRAS.485.5666P} corrected for distance 4.84 kpc and reddening $E(g-y) = 1.827$ mag, are shown with blue and red continuous curves, respectively. The star symbols denote the previously identified massive stars and the bright member stars.
} 
\end{figure}

%
%
%

\section{Results and Analysis} \label{sec3}

\subsection{Stellar clustering/groupings in the S193 complex} \label{group}

To identify clustering/grouping in S193 complex, we estimated the stellar surface density distribution \citep{2009ApJS..184...18G, 2016AJ....151..126S}. 
We used the nearest neighbour (NN) method to generate surface density maps in the region \citep{2005ApJ...632..397G}.  
We took a grid size of 6$^{\prime\prime}$ and determined local surface density by varying radial distance so that it 
encompasses through the nearest 6$^{th}$ NN in 2MASS catalog. 
This technique has been used and described in our previous work also \citep{2020ApJ...891...81P,2022ApJ...926...25P}.
We have shown the density contours in Figure \ref{nirrgb}a with the yellow colour, and the lowest contour is drawn 
at 1$\sigma$ above the mean stellar density (8.7 stars/arcmin$^{2}$), and step size of 1$\sigma$ (2.9 stars/arcmin$^{2}$). 
We can see a fragmented overdensity of stars in the S193 region with three sub-groupings, one at the centre (CC), another at the North-East (NE) direction, and another at the South-East direction (SEC). 
All three sub-groupings are connected through the lowest density contours and seem to be part of a similar population. The clusterings at the centre and NE direction coincide with the previously identified star clusters, [BDS2003]57 and T162, respectively. We have roughly estimated the cluster sizes by visually inspecting the surface density contours. The box enclosing different cluster regions are marked in Figure \ref{nirrgb} with red boxes. In this way, the size of the clustering [BDS2003]57 is found to be $\sim$ (1$\times$1) arcmin$^{2}$ while the size of the Tuesch 162 and SEC is found to be $\sim$ (1.4$\times$1.4) arcmin$^{2}$, (1.5$\times$1.4 arcmin$^{2}$), respectively.
From the 2MASS NIR data, we found 40 stars inside the [BDS2003]57, it has a mean stellar density of $\sim$ 34 stars/arcmin$^{2}$. The Tuesch 162 contains 29 stars inside it with a mean stellar density of $\sim$ 14.1 stars/arcmin$^{2}$, and the SEC has 30 stars inside and it has a mean stellar density of 13.3 stars/arcmin$^{2}$. As the 2MASS data is bit shallow, there is a probability of finding more low-mass/embedded stars within the cluster region, thus above estimates can be taken as lower limits.
Figure \ref{nirrgb} shows the colour composite image of three clusters (see b, c, and d panels) made using $JHK$ band images.
The central clustering [BDS2003]57 (cf. Figure \ref{nirrgb}b) is covered in our TANSPEC observations, while the other two clusterings (cf. Figure \ref{nirrgb} c, d) are covered in our TIRSPEC observations (cf. Section \ref{nird}). The identified groupings ([BDS2003]57, T162, and SEC) lie within a circle centred at $\alpha_{2000}$: 02$^{h}$47$^{m}$34$^{s}$.29, $\delta_{J2000}$: 61$\degr$57$\arcmin$20.93$\arcsec$ and radius of 3$^{\prime}$ {(shown with a white circle in Figure \ref{nirrgb}a).

\subsection{Membership analysis of the stars in the S193 complex} \label{member}

Recently released Gaia data release 3 \citep{2021A&A...649A...1G} is becoming highly impactful in the study of star clusters by providing precise 
measurements of the proper motion (PM) and parallax of stars.
As we have found three inter-connected stellar groupings/clusterings in the S193 complex (Section \ref{group}), it is now important 
to find whether these clusterings are connected in the proper motion space and further estimate their member stars. 
In Figure \ref{vpd}, we have shown the vector point diagrams (VPDs) of three clusterings in which proper motions in RA and DEC are plotted as X and Y axes, respectively. 
In the VPDs, we can see that the stars in all three clusterings show the distribution around 
$\mu_\alpha$cos($\delta$) = -0.70 mas yr$^{-1}$ and $\mu_\delta$ = 0.17 mas yr$^{-1}$, this suggests that the three clusters are also interconnected in the proper motion space.
We thus determined the membership probability of all the stars in a circular FOV of  3$^{\prime}$ (which accommodates all three clusterings, see Section \ref{group}) using the method described in \citet{1998A&AS..133..387B} and \citet{2020MNRAS.498.2309S}.
The procedure starts with determining frequency distributions of cluster stars ($\phi^\nu_c$) and field stars ($\phi^\nu_f$) using equations 3 and 4 of \citet{1998A&AS..133..387B}.
The PM centre and its corresponding dispersion for the member stars ($\mu_\alpha$cos($\delta$) = -0.70 mas yr$^{-1}$, 
$\mu_\delta$ = 0.17 mas yr$^{-1}$, $\sigma_c$, $\sim$0.5 mas yr$^{-1}$) as well as for the field stars 
($\mu_{xf}$ = 0.29 mas yr$^{-1}$, $\mu_{yf}$ = -0.50 mas yr$^{-1}$, $\sigma_{xf}$ = 4.10 mas yr$^{-1}$ and $\sigma_{yf}$ = 2.85 mas yr$^{-1}$) 
is determined using the procedure previously used and described in \citet{2020ApJ...891...81P, 2022ApJ...926...25P} and \citet{2020MNRAS.498.2309S}.
We calculated PM dispersion for the cluster by using the radial velocity dispersion of 1 kms$^{-1}$ for open clusters \citep{1989AJ.....98..227G} and assuming a distance of 4.84 kpc (present estimate), the PM dispersion comes out to be 0.5 mas yr$^{-1}$. We used this value as a PM dispersion for the cluster stars (for both RA and DEC), the same method has been used in previous studies \citep{2013MNRAS.430.3350Y,2020MNRAS.498.2309S,2022ApJ...926...25P}. The field stars are taken from the VPD of all stars within a 3' circle enclosing all three clustering and roughly falling outside of the PM distribution of cluster stars (for more details, see \citet{2020MNRAS.498.2309S}).
Finally, we have determined the membership probability, i.e., the ratio of the distribution of cluster stars to all stars, using the below-mentioned equation.
 
\begin{equation}
P_\mu(i) = {{n_c\times\phi^\nu_c(i)}\over{n_c\times\phi^\nu_c(i)+n_f\times\phi^\nu_f(i)}}
\end{equation}

Where $n_c$ (=0.19) and $n_f$ (=0.81) correspond to the normalized number of stars for the cluster and field region, respectively.

In Figure \ref{pm}, membership probability, errors in proper motion, and parallaxes are plotted against the G magnitude. 
By applying the membership criteria of considering only those stars as a member which have a membership 
probability P$_\mu$ $>$ 80 \%, we identified 80 stars as the member of clustering in the S193 complex. Figure \ref{pm} shows the member stars as green circles.
Figure \ref{pm} depicts that the member stars are quite effectively separated from the field stars in 
the brighter end while in the fainter end, they are not well separated (see top panel of Figure \ref{pm}) because of 
the large uncertainties in the proper motion at fainter magnitudes (middle panel). 
In Table \ref{PMT}, we have tabulated the 80 member stars of the S193 complex.

\subsection{Reddening and distance of the S193 complex} \label{dist}

Few distance estimates are available in the literature for the S193 complex but with different values.
\citet{2007A&A...470..161R} reported a distance of $2.4\pm0.32$ kpc based on the individual distances of S192 and S193 H\,{\sc ii} regions, while, \citet{2008A&A...484..361Q} and \citet{2011AJ....141..123A} reported distances as 5.2 kpc and $2.96\pm0.54$ kpc, respectively. \citet{2008A&A...484..361Q} reports a distance based on the galactic rotation curve and the accompanying cloud's CO velocity, whereas \citet{2007A&A...470..161R} and \citet{2011AJ....141..123A} report spectrophotometric distances.\\
Out of the 80 member stars of the S193 complex, the distance of the 23 stars
having parallax values with good accuracy (i.e., error$<$ 0.1 mas, red triangles in Figure \ref{pm})}
having distance estimation in \citet{2021AJ....161..147B}, is further used to constrain the distance of the clusters and the
S193 complex. 
We estimated the distance of each clustering in the S193 complex by taking the mean of the distance of the above 
stars provided in \citet{2021AJ....161..147B}, which belongs to that particular cluster. 
In this way, the distance of the cluster [BDS2003]57 is found to be 4.59$\pm$1.4 kpc while the 
distance of the T162 and SEC is found to be 4.96$\pm$0.67 kpc and 4.51$\pm$1.6 kpc, respectively. 
The distance of the whole S193 complex is found to be 4.84$\pm$1.4 kpc.
The distance of the clusters matches each other and the distance of the S193 complex within the error, which suggests that all the clustering are 
at the same distance and are part of the S193 complex. Our derived distance of the S193 complex matches with the distance 
derived by \citet{2008A&A...484..361Q} and \citet{1995AJ....109.2611C}, while differs with the distance 
estimated by \citet{2007A&A...470..161R} and \citet{2011AJ....141..123A}. We have also calculated the distance of massive stars in the complex 
using distance estimates from \citet{2021AJ....161..147B}. The distance of the massive star S192-1 is 5.99$\pm$0.45 kpc, 
while the distance of the massive star S193-1 and S193-3 is found to be 5.06$\pm$0.64 and 10$\pm$2 kpc, respectively. 
The distance of the massive stars S192-1 and S193-1 matches the distance of the S193 complex within error, while the S193-3 is much far away. \citet{2007A&A...470..161R} also estimated the distance of the massive star S193-3 as 8.0 kpc and concluded that the S193-3 is a background star and not a part of the S193 complex.
 We have also used the CMD of the stars in the S193 complex to validate the derived distance of the S193 further. This technique is very much 
reliable and widely used by many authors \citep[for example,][]{1994ApJS...90...31P,2006AJ....132.1669S, 2017MNRAS.467.2943S, 2020MNRAS.498.2309S}. 
In Figure \ref{ccd}, we have shown the $(g-y)$ vs. $g$ CMD of the stars in the S193 complex (black dots) taken from the PS1 archive (see Table \ref{archilog}). 
The member stars in Figure \ref{ccd} are shown with green circles.
The reddening value in the direction of the S193 is taken from the Bayester dust map by \citet{2019ApJ...887...93G}. We retrieved the E(g-y)=1.827 mag for our studied target from the extinction map of \citet{2019ApJ...887...93G}, then it is converted to Av using the color relations given by \citet{2019ApJ...877..116W}. We found the reddening value to be A$_{V}$= 2.41 mag in the direction of the S193 complex. The ZAMS (continuous blue curve) and PMS isochrone of 1.0 Myr (continuous red curve) is also shown in the figure, which is taken from \citet{2019MNRAS.485.5666P}. 
Both the ZAMS and isochrone are corrected for the reddening (E(g-y)= 1.827 mag) and distance (4.84 kpc). 
The ZAMS matches the distribution of the member stars, which gives 
confirmation regarding the distance and reddening of the S193 complex.

Apart from the previously identified massive stars (S192-1, S193-1, and S193-3), the S193 complex hosts 
few other bright member stars which are shown with star symbols in Figure \ref{ccd}.
The star BG2 lies at the boundary of the S193 complex and has no radio or 22 $\mu$m emissions 
traced around it (cf. Figure \ref{fig1}). This star is located at a  distance of 2.75$\pm$0.21 kpc \citep{2021AJ....161..147B}, 
thus is probably a foreground star.
The star BG1 and S194-1 lie inside the H\,{\sc ii} region S193 and S194, respectively, and seem to be 
associated with the radio emission with an envelope kind of structure visible in the WISE 22 $\mu$m band  (cf. Figure \ref{fig1}). 
The location of the  BG1 and S194-1 in the CMD indicate their spectral types 
as B5V and B1.5V, respectively. The star S192-1 (previously B2.5V star) seems to be around B1.5V, while S193-1 (previously B2.5V) seems to be around B2.5V in the CMD. With our radio analysis (cf. Section  \ref{ioni}), we found that the ionising source responsible for creating the H\,{\sc ii} regions S192, S193, and S194, should have a spectral type between B0-B0.5. Hence, the BG1 (B5V) along with the S193-1 (B2.5V) may be jointly responsible for creating the H\,{\sc ii} region S193. The brightest massive stars S192-1 (B1.5V) and S194-1 (B1.5V) could be the ionizing source of S192 and S194, respectively.




\begin{figure*}
\centering
\includegraphics[width=0.48\textwidth]{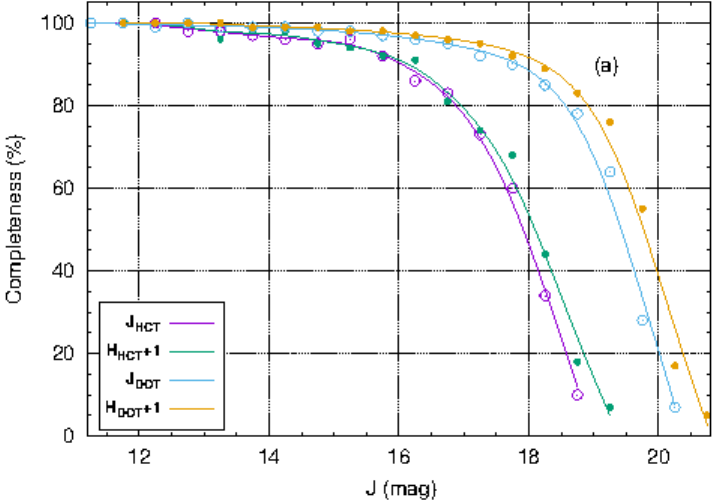}
\includegraphics[width=0.48\textwidth]{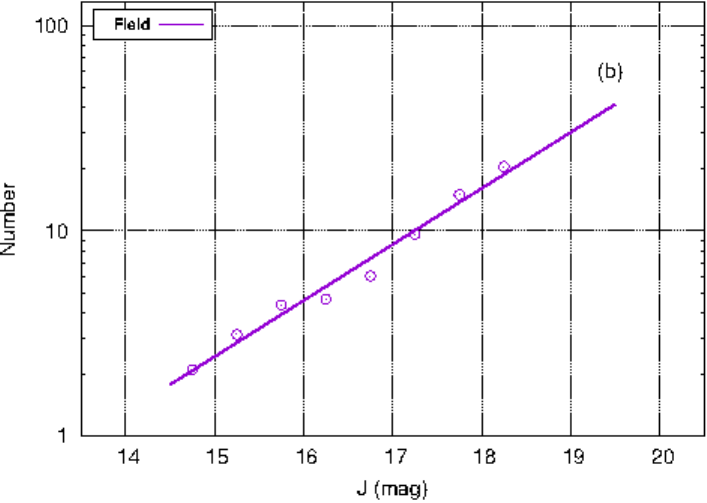}
\caption{\label{cft} (a) Completeness factor as a function of $J$ magnitude
derived from the artificial star experiments ({\it ADDSTAR}, see Section \ref{mf1} for details)
on the TANSPEC, TIRSPEC $J$ and $H$ band images.
The $H$-band completeness factor is off-setted by the mean colour of the MS stars (i.e., 1.0 mag).
The continuous curves are the smoothed bezier curves for the data points for completeness.
(b) Field stars distribution generated by using a nearby reference
field (magenta circles) using TIRSPEC $J$ band data. A straight line is a least square fit to the data points.}
\end{figure*}

\begin{figure*}
\centering
\includegraphics[width=0.48\textwidth]{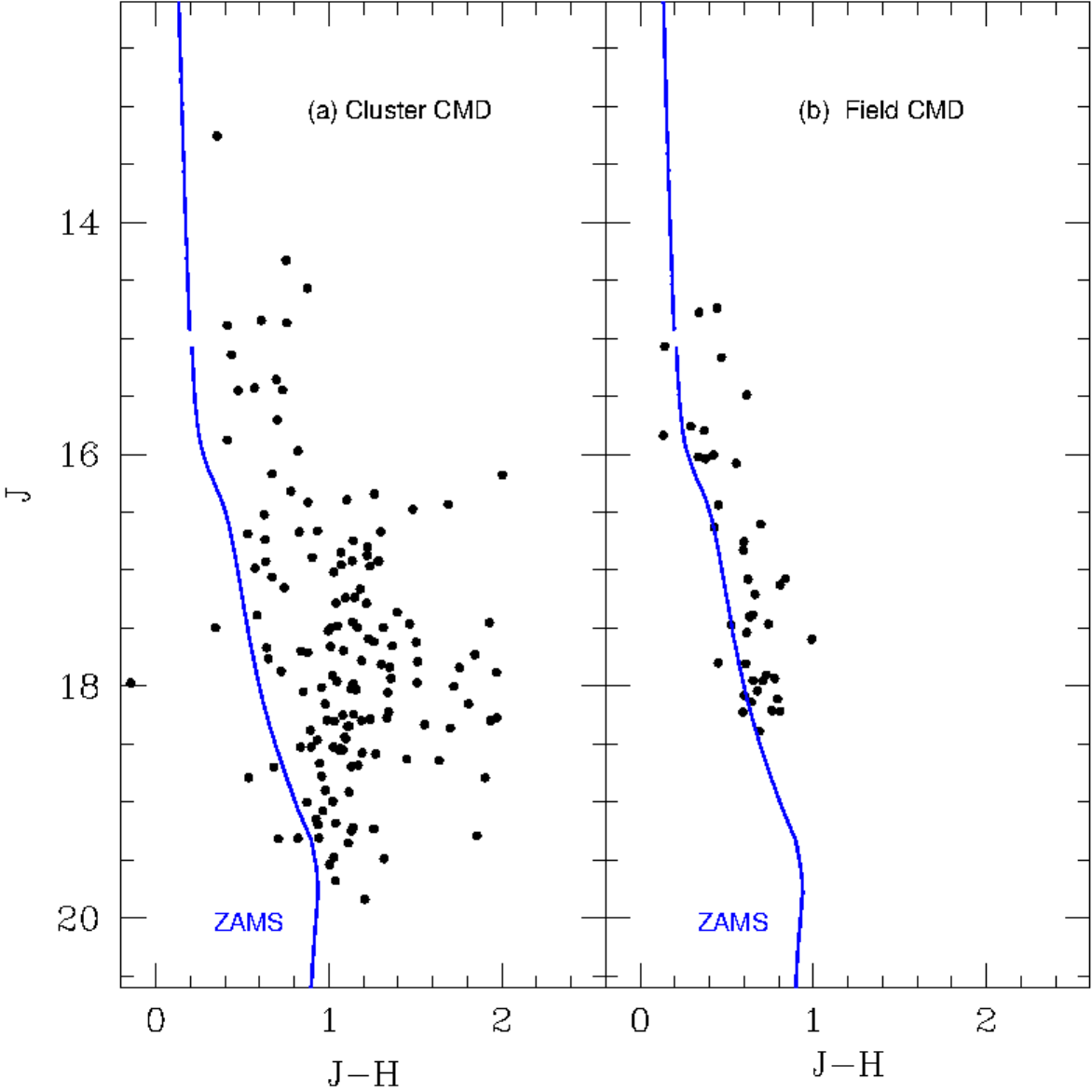}
\includegraphics[width=0.48\textwidth]{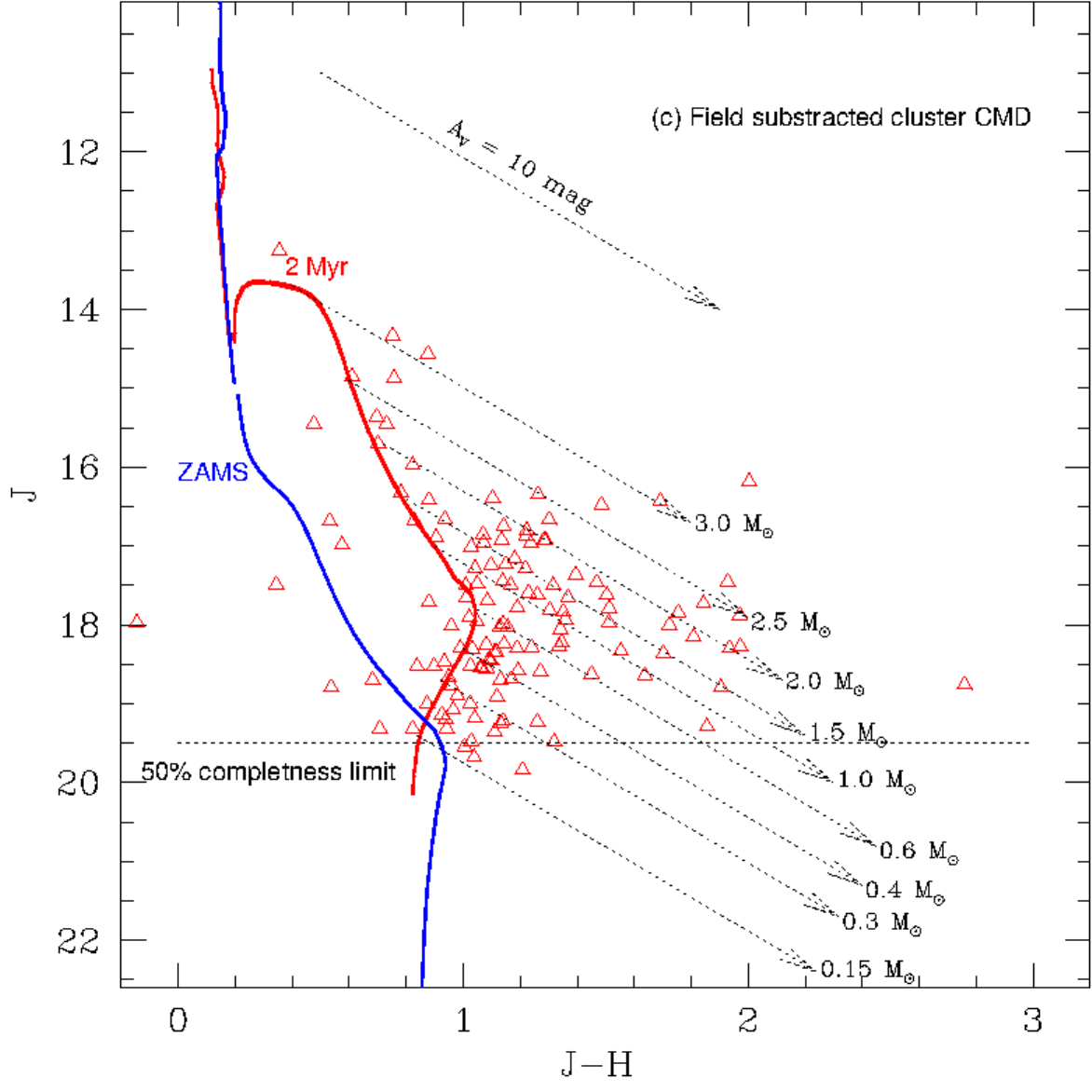}
\caption{\label{band} $J$ vs. $J-H$ CMD for (a) stars in the [BDS2003]57 cluster region, (b) stars in the reference region, and
(c) statistically cleaned sample of stars in the [BDS2003]57 cluster region.
The ZAMS (blue curve) and isochrone of 2 Myr (red curve) by \citet[][]{2019MNRAS.485.5666P}
along with reddening vector for different mass stars (black slanted dashed lines in panel (c)) are also shown.
Stars which are on the left side of the 2 Myr isochrone in panel (c)
are used to estimate the MF of the region.
All the curves are corrected for the  distance of 4.84 kpc and foreground extinction of A$_V$=2.41 mag.
The black horizontal dashed-line in panel (c) represents the 50\% completeness limit of the TANSPEC data.}
\end{figure*}

\begin{figure}
\centering
\includegraphics[width=0.48\textwidth]{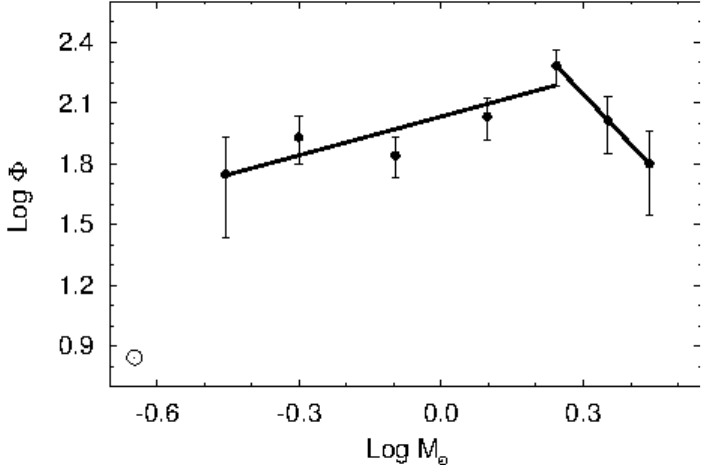}
\caption{\label{mf} 
A plot of the MF for the statistically cleaned CMD for the stellar sources in the  [BDS2003]57 cluster region.
Log $\phi$ represents log($N$/dlog $m$). The error bars represent $\pm\sqrt N$ errors. The solid line shows the least
squares fit to the MF distribution (black dots).
The open circle is the data point falling below the completeness limit of 0.3 M$_\odot$. 
}
\end{figure}

\subsection{Mass Function} \label{mf1}
The mass function (MF) is an important statistical tool for understanding the formation of stars and is
expressed by a power law,
$N (\log m) \propto m^{\Gamma}$ and  the slope of the MF is given as:
$ \Gamma = d \log N (\log m)/d \log m  $,
where $N (\log m)$ is the number of stars per unit logarithmic mass interval.
We have used our deep NIR data from TANSPEC observations to generate the MF of the central cluster [BDS2003]57
which has the maximum number density among all three clusterings found in this region.
We have utilized the NIR  $J$ versus $J-H$ CMDs of the
sources in the target region and that of the nearby field region of equal area
and decontaminated the former sources from the foreground/background
stars and corrected for data in-completeness using a statistical subtraction method
\citep[cf.][]{2022ApJ...926...25P, 2020ApJ...891...81P, 2020MNRAS.498.2309S, 2017MNRAS.467.2943S, 2012PASJ...64..107S, 2008AJ....135.1934S}.
We have calculated the completeness factor (CF) for both $J$ and $H$ bands and taken the minimum of them according to the mean $J-H$ colours ($\sim$ 1 mag) of the MS stars
as our final completeness value \citep[see][]{1991A&A...250..324S}. To detemiine the CF, we used the IRAF {\it ADDSTAR} routine as described in our previous publications \citep{2022ApJ...926...25P, 2020ApJ...891...81P}. We have derived the CF for the TANSPEC and TIRSPEC data, which is shown as a function of $J$-band in the left panel of Figure \ref{cft}.
As expected, the incompleteness of the data increases with increasing magnitude and the 
photometric data is more than 50\% complete up to $\sim$19.5 mag and  $\sim$18.0 mag for the TANSPEC and TIRSPEC observations, respectively.
As TIRSPEC observations are shallow and we do not have field region observations using the TANSPEC, we have used the distribution of 
a nearby reference field in TIRSPEC pointing and the completeness corrected luminosity function (right panel of Figure \ref{cft}) of this region is used to 
calculate the number of field stars between $J\sim$ 18.0 to 19.5 mags.

In  Figure \ref{band}, we have shown the $J$ versus $J-H$ CMDs for the
stars lying within the central cluster `[BDS2003]57' in panel (a) and for those in the reference field region in panel (b).
In panel (c), we have plotted the statistically cleaned $J$ versus $J-H$
CMD for the central cluster `[BDS2003]57' of $\sim$2 Myr of age.
The number of stars in different mass bins was then calculated by counting the stars along the reddening vector for a 2 Myr isochrone. 
The corresponding MF is plotted in Figure \ref{band} (right panel).
The 50\% completeness limit (J $\sim$9.5 mag) of the TANSPEC photometry 
corresponds to the detection limit of $\sim$ 0.3 M$_\odot$ stars
of an age of $\simeq$2 Myr age, embedded in the nebulosity up to $A_V\simeq$2.4 mag (foreground reddening) and $\sim$3 mag (differential reddening) (cf.  Figure \ref{band} (left panel) `c').
There seems to be a break in the slope of the MF distribution of `[BDS2003]57 cluster 
at $\sim$1.5 M$_\odot$ and the slope `$\Gamma$' in the mass range  $3>M_\odot>1.5$ and  $1.5>M_\odot>0.3$ 
is estimated as $-2.46\pm0.01$ and $+0.64\pm0.20$, respectively (cf. Figure \ref{mf}).

\subsection{YSOs identification and classification} \label{yid}
\begin{figure*}
\centering
\includegraphics[width=0.8\textwidth]{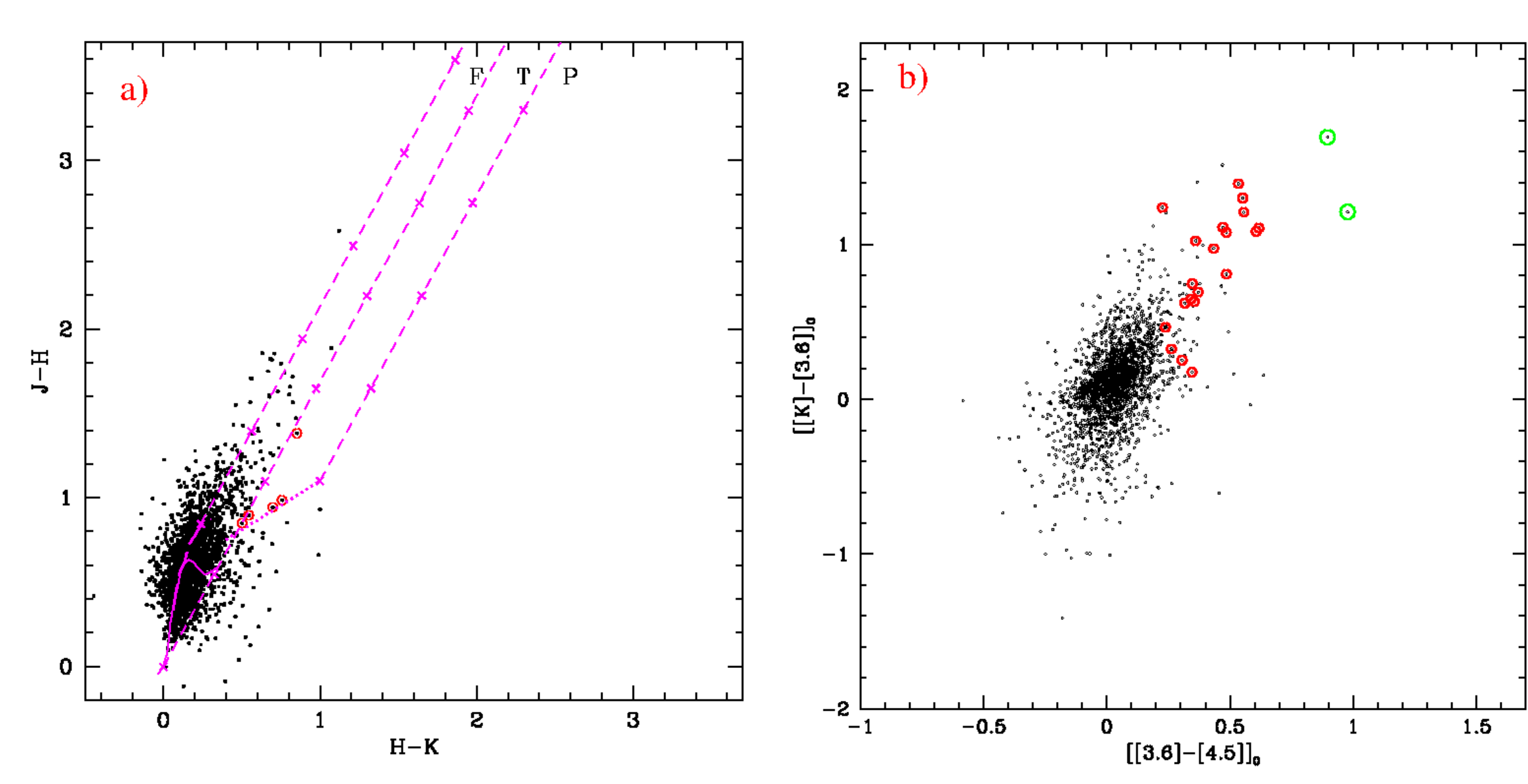}
\caption{\label{yso} a) ${[H-K]}$ vs. ${[J - H]}$ TCD for the sources corresponding to the same FOV. We show the reddened MS and giant branches \citep{1988PASP..100.1134B} by thick magenta dashed curves in the figure. The locus of dereddened CTTSs \citep{1997AJ....114..288M} is shown by dotted magenta line. We also show parallel magenta dashed lines drawn from the tip (spectral type M4) of the giant branch (left reddening line), from the base (spectral type A0) of the MS branch (middle reddening line) and from the tip of the intrinsic CTTS line (right reddening line). The increment of  $A_V$ = 5 mag on the reddening lines is shown with crosses, the red circles are the identified Class\,{\sc ii} sources. b) ${[[3.6] - [4.5]]_0}$ vs. ${[[K] - [3.6]]_0}$ TCD of all the sources inside the FOV of S193 complex ($\sim10^\prime\times 10^\prime$). We used the colour criteria by \citet{2009ApJS..184...18G} to identify and classify the YSOs. We show the identified Class\,{\sc i} and Class\,{\sc ii} YSOs with green and red circles, respectively.}
\end{figure*}

We have used combined NIR catalogue (see Section \ref{sec2}) along with $Spitzer$ MIR data to identify the YSOs in the $\sim10^\prime\times 10^\prime$ FOV of the S193 complex.
The classification schemes of identifying YSOs are discussed and explained in our previous publications \citep{2020ApJ...891...81P, 2022ApJ...926...25P}.
In Figure \ref{yso} (a) we show the NIR TCD (${[H-K]}$ vs. ${[J - H]}$) made by plotting the stars in our combined NIR catalogue (black dots), the thick magenta curve in the figure shows the redenned MS and giant branches, dotted magenta line shows the locus of dereddened CTTS, and dashed magenta lines show the redenning lines. The classification scheme by \citep{2009ApJS..184...18G} is used to identify the YSOs using NIR TCD.
Figure \ref{yso} (b) shows the ${[[3.6] - [4.5]]_0}$ vs. ${[K]-[3.6]]_0}$ TCD, made using the $Spitzer$ data, we adopted the method by \citet{2009ApJS..184...18G} to identify and classify the YSOs.
Finally, we were able to identify 2 Class\,{\sc i} and 25 class Class\,{\sc ii} YSOs using MIR data, and 5 Class\,{\sc ii} sources using NIR data.


\begin{figure*}
\centering
\includegraphics[width=0.95\textwidth]{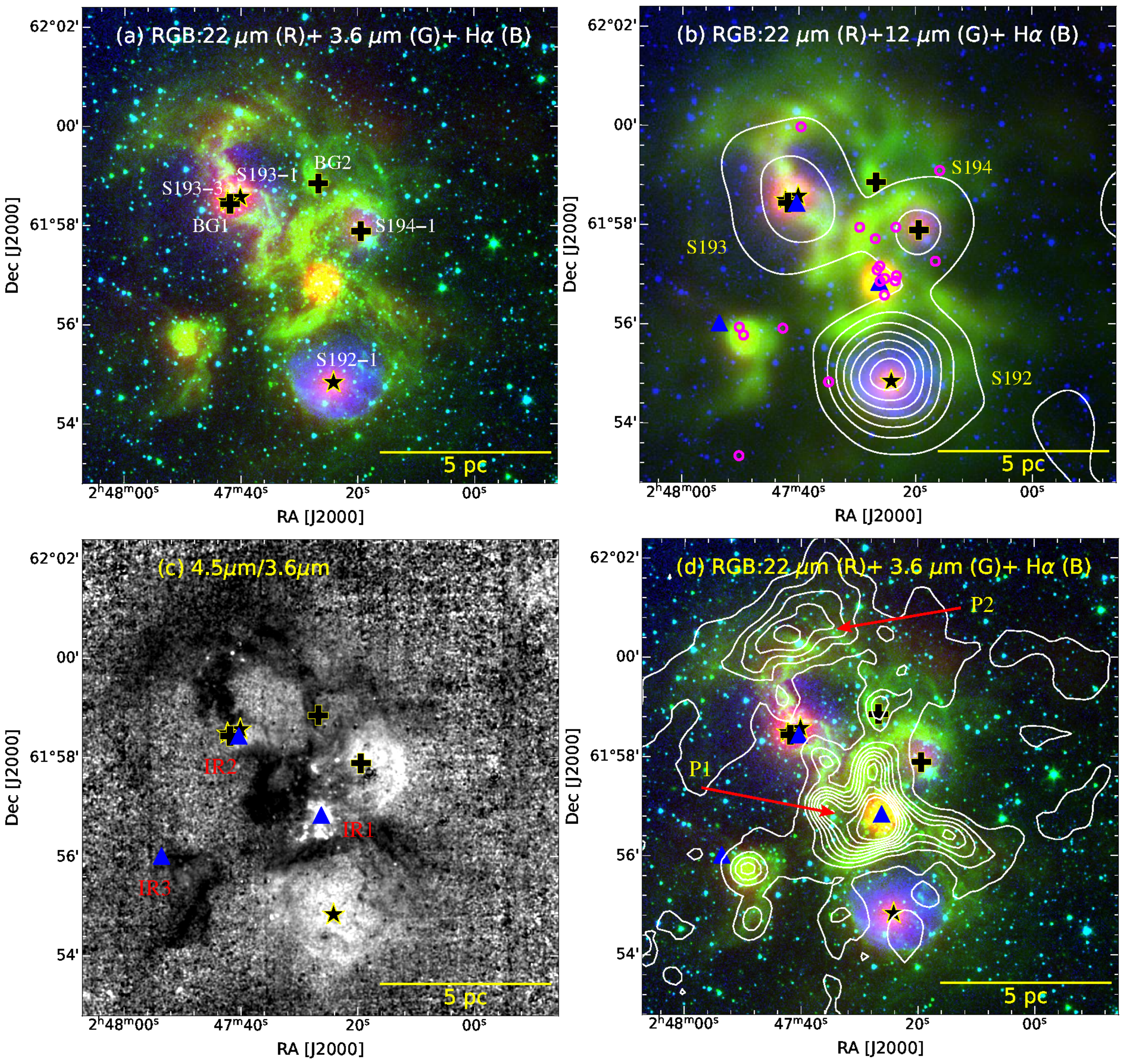}
\caption{\label{fig1} (a) Colour composite image of the S193 complex ($\sim10^\prime\times 10^\prime$ FOV) made using $WISE$ 22 $\mu$m (red),
$Spitzer$ 3.6 $\mu$m (green) and $IPHAS$ H$\alpha$ (blue. (b) Colour composite image of the S193 complex ($\sim10^\prime\times 10^\prime$ FOV) made using $WISE$ 22 $\mu$m (red), $WISE$ 12 $\mu$m (green),
$IPHAS$ H$\alpha$ (blue) images is overplotted with CGPS radio continuum contours. The individual H\,{\sc ii} regions (S192, S193, and S194) are also marked over the image, and the magenta circles show the YSOs identified in the present study. The lowest radio contour is at 5 K and the step size is 0. 5 K. (c) $Spitzer$ ratio map of the S193 complex obtained by dividing $Spitzer$ 4.5 $\mu$m image by $Spitzer$ 3.6 $\mu$m image. (d) The figure shows the same image as in panel (a), overplotted with the $JCMT$ ($^{12}$CO(J =3$-$2)) contours. The lowest contour is at 2 K km s$^{-1}$ and the step size is 4 K km s$^{-1}$. The arrows marked two prominent peaks (P1 and P2) in CO emission. In the images, star symbols represent the previously identified massive stars, while triangles are the IRAS sources. We have also shown previously identified Be star with the cross symbol and the candidate massive sources with the plus symbol.
}
\end{figure*}

\subsection{Multiwavelength view of the S193 complex} 

\subsubsection{Distribution of the dust} \label{dust}

In Figure \ref{fig1}a, we have shown the color composite image of the S193 complex ($\sim10^\prime\times 10^\prime$ FOV) made by
using the $WISE$ 22 $\mu$m (red), $Spitzer$ 3.6 $\mu$m (green), and $IPHAS$ H$\alpha$ (blue) images. The previously identified massive stars in the S193 complex (see Section \ref{sec1}) 
are shown with the star symbols, while a Be type star and candidate massive sources (see Section \ref{sec1}) 
are shown with the cross and plus symbols, respectively. The H\,{\sc ii} regions in the S193 complex (S192, S193, and S194, marked in Figure \ref{fig1}b) are well traced by 
the H$\alpha$ emission in the region.
We can also see the warm dust envelopes around the massive stars and central cluster ([BDS2003]57) traced by the $WISE$ 22 $\mu$m emission. 
Prominent poly-cyclic aromatic hydrocarbon (PAH) features at 3.3, and 11.3 $\mu$m are indicative of PDRs (photo-dissociation regions) which lies within the bandwidth of the 3.6 $\mu$m and 12 $\mu$m images \citep[see e.g.,][]{2004ApJ...613..986P}.
The PDR regions traced by the 3.6 $\mu$m emission are distributed around the H\,{\sc ii} regions in arc-type or circular sub-structures. 
The S192 H\,{\sc ii} region, having PDRs circularly distributed around the warm gas with massive stars at the center,
looks like  a classical example of a MIR bubble \citep[cf.][]{2006ApJ...649..759C}.

In Figure \ref{fig1}b, we have shown the color composite image of the S193 complex made by using the $WISE$ 22 $\mu$m (red), $WISE$ 12 $\mu$m (green), and $IPHAS$ H$\alpha$ (blue) images. 
The image is also over-plotted with the radio continuum contours from the $CGPS$ 1420 MHz map. The spatial distribution of the young stellar objects (YSOs; see Section \ref{yid}) also shown with the magenta color circles.
The heated (22 $\mu$m emission) and the ionised gas (radio emission) seem to be distributed at the location of the massive stars and are surrounded
by the PDRs (12 $\mu$m emission). The YSOs in this complex are generally located, away from the massive stars, in regions with higher gas and dust emissions (12 $\mu$m emission).
We have also shown the $Spitzer$ ratio map (4.5 $\mu$m/3.6 $\mu$m) in the Figure \ref{fig1}c. 
The prominent molecular hydrogen ($H_2$) line emission ($\nu$ = 0--0 $S$(9); 4.693 $\mu$m) and the Br-$\alpha$ emission (at 4.05 $\mu$m) 
comes within the bandwidth of $Spitzer$ 4.5 $\mu$m band. 
Therefore, the bright and dark regions in this ratio map trace mainly the  Br-$\alpha$ emission and PDRs, respectively \citep[for more details, see][]{2020ApJ...891...81P}.
Note that both of these {\it Spitzer} images have the same PSF, allowing to remove of point-like sources
as well as continuum emission \citep[for more details, see][]{2017ApJ...834...22D}.
In Figure \ref{fig1}c, the bright regions representing the $H_2$ and Br-$\alpha$ emission almost mimic the radio continuum emission from the hot ionised gases.
The distribution of the dark lanes (PDRs) around the bright regions/radio emission  
indicates the impact of massive stars in the S193 complex. In Figure \ref{fig1}d, we show the same image as in \textbf{Figure \ref{fig1}a}, also overplotted with the $JCMT$ ($^{12}$CO(J =3$-$2)) contours (see Section \ref{jcmK}). Interestingly, we can also see a bright patch at P1 (see Figure \ref{fig1}c) located away from the radio continuum emission. The cavities in the molecular cloud are clearly visible which shows the impact the H\,{\sc ii} regions have produced in their surrounding environment. The CO emission seems to peak at [BDS2003]57 (P1), above the H\,{\sc ii} region S193 (P2), and at the SEC clustering.
  
\subsubsection{Distribution of the ionised gas} \label{ioni}
To trace the distribution of ionised gas in the S193 complex, we have used both the radio continuum (free-free emission) and the H$\alpha$
images. In Figure \ref{fig1}b, we can see that the H$\alpha$ emission in the S193 complex is mostly associated
with the H\,{\sc ii} regions. In the H\,{\sc ii} region S192, the H$\alpha$ emission is almost circularly enclosed by the 
PDRs like a MIR bubble. In the case of H\,{\sc ii} regions S193 and S194, the  H$\alpha$ emission  is more or less extended and is partially surrounded by the
PDRs in arc-like sub-structures. Radio continuum contours from the 1.42 GHz map of $CGPS$ are also overplotted in Figure \ref{fig1}b.
They also show similar distribution with separate peaks for each of the H\,{\sc ii} regions. 
The massive star S192-1 (located inside the H\,{\sc ii} region S192) and S193-1 (located inside the H\,{\sc ii} region S193) 
might be the ionizing sources of the S192 and S193  H\,{\sc ii} regions, respectively.
There is no previously identified massive star inside the boundary of H\,{\sc ii}
region S194, but there is a bright star showing an envelope of warm dust at 22 $\mu$m (plus symbol), which could be the possible ionizing source of S194. 
 
By estimating the Lyman continuum flux associated with the ionised gas in each of the H\,{\sc ii} regions, we can also verify the spectral type of the ionizing sources.
Thus, this could lead us to constrain the massive star responsible for creating the H\,{\sc ii} regions S192, S193, and S194.
We have used the following equation from \citet{2016A&A...588A.143S} to determine the Lyman continuum flux associated with each H\,{\sc ii} region.

\begin{equation}
\begin{split}
N_\mathrm{UV} (s^{-1})& = 7.5\, \times\, 10^{46}\, \left(\frac{S_\mathrm(\nu)}{\mathrm{Jy}}\right)\left(\frac{D}{\mathrm{kpc}}\right)^{2} \\
&\left(\frac{T_{e}}{10^{4}\mathrm{K}}\right)^{-0.45} \times\,\left(\frac{\nu}{\mathrm{GHz}}\right)^{0.1}
\end{split}
\end{equation} 
 
In the above equation, N$_{UV}$ denotes the Lyman continuum photons per second, T$_e$ denotes the electron temperature, $\nu$ is 
the frequency, S$_\nu$ is the integrated flux, D is the distance of the S193 complex, i.e., 4.84 kpc. We have assumed that all ionization flux in
an H\,{\sc ii} region is generated by a single massive star and adopted the value of $T_e$ as 10000 K. To calculate the integrated 
flux, we have used the task {\sc jmfit} in AIPS on the  NVSS (1.4 GHz) map. 
The values of the integrated flux for the H\,{\sc ii} regions S192, S193 and S194 are 67 mJy ($\sigma$=0.63 mJy/beam), 
42.9 mJy ($\sigma$=0.54 mJy/beam), and 11 mJy ($\sigma$=0.44 mJy/beam), respectively. 
The radius of the H\,{\sc ii} regions S192, S193, and S194 are found to be 1.85 pc, 1.79 pc, and 1.20 pc, respectively, at a distance of 4.84 kpc. 
The log(N$_{UV}$) values for the H\,{\sc ii} regions S192, S193, and S194 are then estimated as 47.08, 46.89, and 46.30, respectively. 
By comparing these log(N$_{UV}$) values with those given in \citet{1973AJ.....78..929P}, we have found that
the spectral type of the ionizing sources for all the  H\,{\sc ii} regions in the S193 complex is between B0.5-B0 type.

We have also calculated the dynamical age of the H\,{\sc ii} regions by using the following equation by \citet{1980Natur.287..373D}:

\begin{equation}
t_{dyn} = \Big(\frac{4R_s}{7c_s}\Big) [\Big(\frac{R_{H\,{\sc ii}}}{R_s}\Big)^{7/4}-1]
\end{equation}

where, c$_s$ is the isothermal sound velocity in the ionised gas (c$_s$ = 11 km s$^{-1}$) \citep{2005fost.book.....S}, 
R$_{H\,{\sc ii}}$ is the radius of the H\,{\sc ii} region, and R$_s$ is the  Str\"{o}mgren radius of the H\,{\sc ii} region which is  given by:

\begin{equation}
R_s = \Big(\frac{3S_{\nu}}{4\pi{{n_0}^2}{\beta_2}}\Big)^{1/3}
\end{equation}

where, n$_0$ is the initial ambient density (in cm$^{-3}$) and $\beta$$_2$ 
is the total recombination coefficient to the first excited state of hydrogen $\beta$$_2$ = $2.6\times10^{-13}$ \citep{2005fost.book.....S}.
The dynamical age of the H\,{\sc ii} regions S192, S193, and S194 for n$_0$=10$^3$(10$^4$) cm$^{-3}$ are estimated as 0.6 (1.9) Myr, 0.6 (2.0) Myr, 0.4 (1.4) Myr, respectively.





\subsubsection{Embedded $Herschel$ clumps/condensation in the S193 complex} \label{clump}

In this section, we have examined the $Herschel$ column density ($N(\mathrm H_2)$) and temperature map and identified
the embedded structures in the S193 complex. We have used the maps produced for the {\it EU-funded ViaLactea project}
\citep{2010PASP..122..314M} and are publicly available\footnote{http://www.astro.cardiff.ac.uk/research/ViaLactea/}. 
A Bayesian PPMAP approach is used on the $Herschel$ images  at
70, 160, 250, 350 and 500 $\mu$m to produce these maps \citep{2010A&A...518L.100M,2015MNRAS.454.4282M, 2017MNRAS.471.2730M}.

\begin{figure*}
\centering`	
\includegraphics[width=1\textwidth]{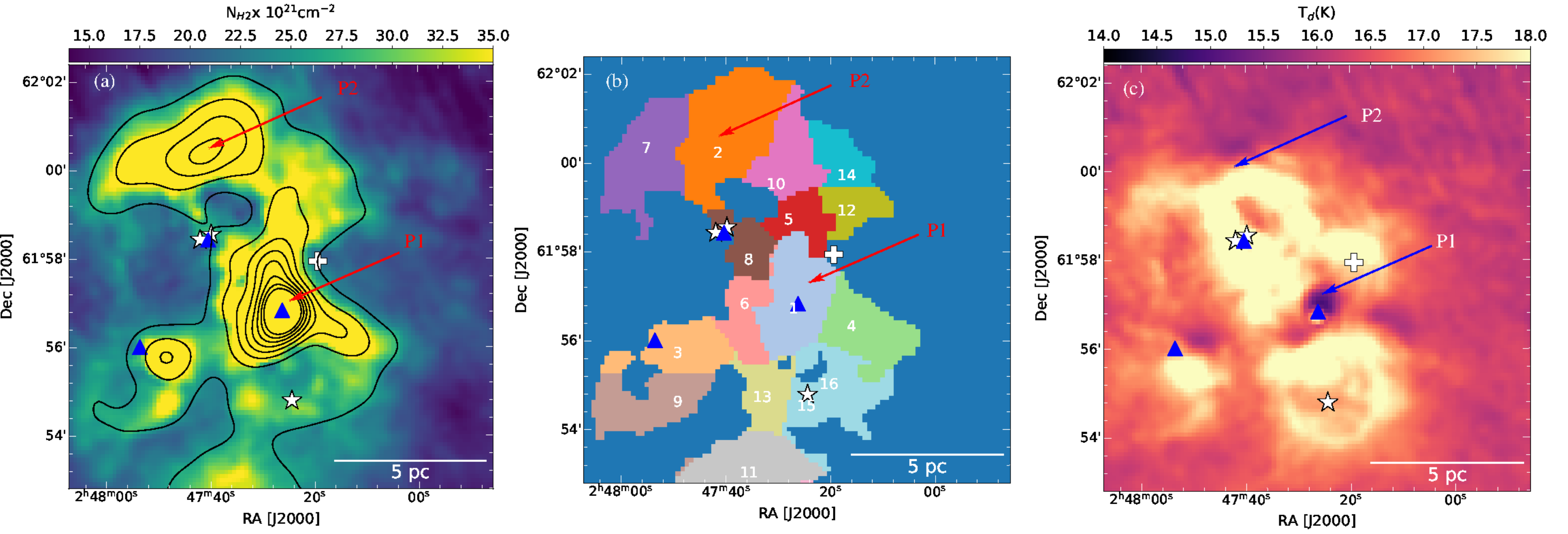}
\caption{\label{fig2} (a) $Herschel$ column density map of the S193 complex overplotted with the column density contours. (b) Clumps identified in the S193 complex using the $Herschel$ column density map are marked with the corresponding IDs. (c) $Herschel$ temperature map, the cold gas clump at position P1 is also marked. 
The position of clumps P1 and P2 is marked in all three panels. The symbols in all three panels are similar to those shown in Figure \ref{fig1}. }
\end{figure*}

We show the $Herschel$ column density map of the S193 complex in Figure \ref{fig2}a. To identify clumps/condensation, 
we have employed the {\sc clumpfind} algorithm \citep{1994ApJ...428..693W} on the $Herschel$ column density map. The {\sc clumpfind} provides the total column density and the corresponding area of each clump as an output.
We identified 16 clumps in our target field, shown in Figure \ref{fig2}b. We have further calculated the mass of each 
of the clump by employing the following equation: 

\begin{equation}
M_{area} = \mu_{H_2} m_H Area_{pix} \Sigma N(H_2)
\end{equation}

where $\mu_{H_2}$ is the mean molecular weight per hydrogen molecule (i.e., 2.8), $m_H$ is the mass of the hydrogen atom, and $Area_{pix}$ 
is the area subtended by one pixel {(i.e., 6$\arcsec$pixel$^{-1}$)}. 
$\Sigma N(\mathrm H_2)$ signifies the total column density estimated using  the {\sc clumpfind} \citep[see also][]{2017ApJ...834...22D}.

The mass and size of the identified clumps are provided in the Table \ref{cd1}. We have the mass range of 224.6 M$_{\odot}$ to 1141.9 M$_{\odot}$ for the clumps identified in the present study. The most massive clump (M$_{\odot}$=1141.9, ID=1)
lies almost at the center of S193 complex coinciding with the boundary of [BDS2003]57 and peak P1 (see Section \ref{dust}). 
Other two massive clumps (M$_{\odot}$=912.2, ID=2) and (M$_{\odot}$=527.8, ID=7) are located near P2 (see Section \ref{dust}). 
We can also see the two clumps (IDs=3 and 9) associated with the SEC clustering (see Section \ref{group}).
{
In Figure \ref{fig2}c, we also show the $Herschel$ temperature map of the S193 complex. The S193 complex can be seen glowing with a temperature greater than 17.5 K in the $Herschel$ temperature map. A temperature around 14 K can be seen at P1, confirming the presence of a cold dust clump at this position.

\begin{figure}
\centering
\includegraphics[width=0.45\textwidth]{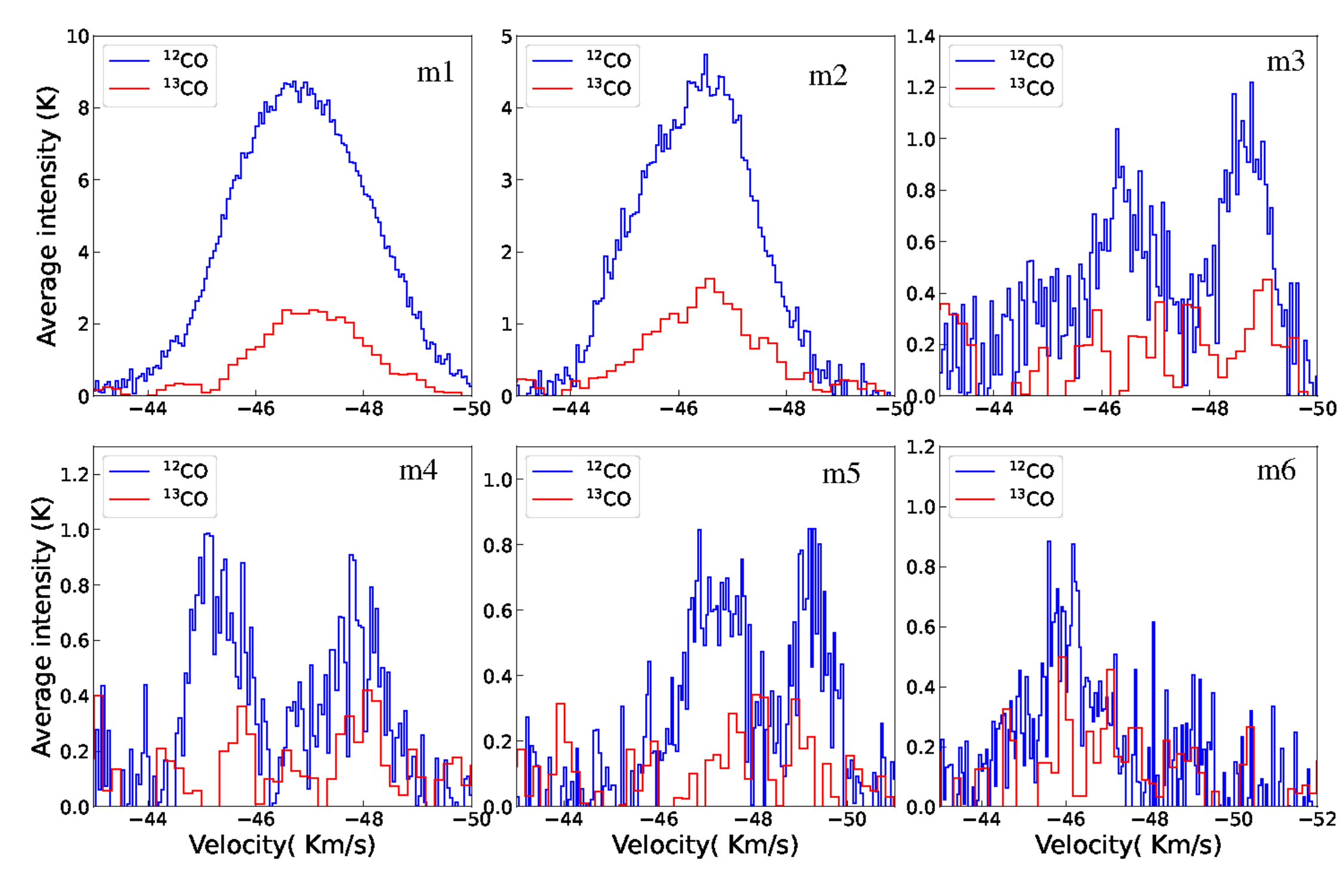}
\caption{\label{spec} The $^{12}$CO(J =3$-$2) and $^{13}$CO(J =1$-$0) profile in the direction of 6 small regions (i.e. m1 to m6;
see Figure \ref{momap}a).
}
\end{figure}

\begin{figure*}
\centering
\includegraphics[height=18cm, width=18cm]{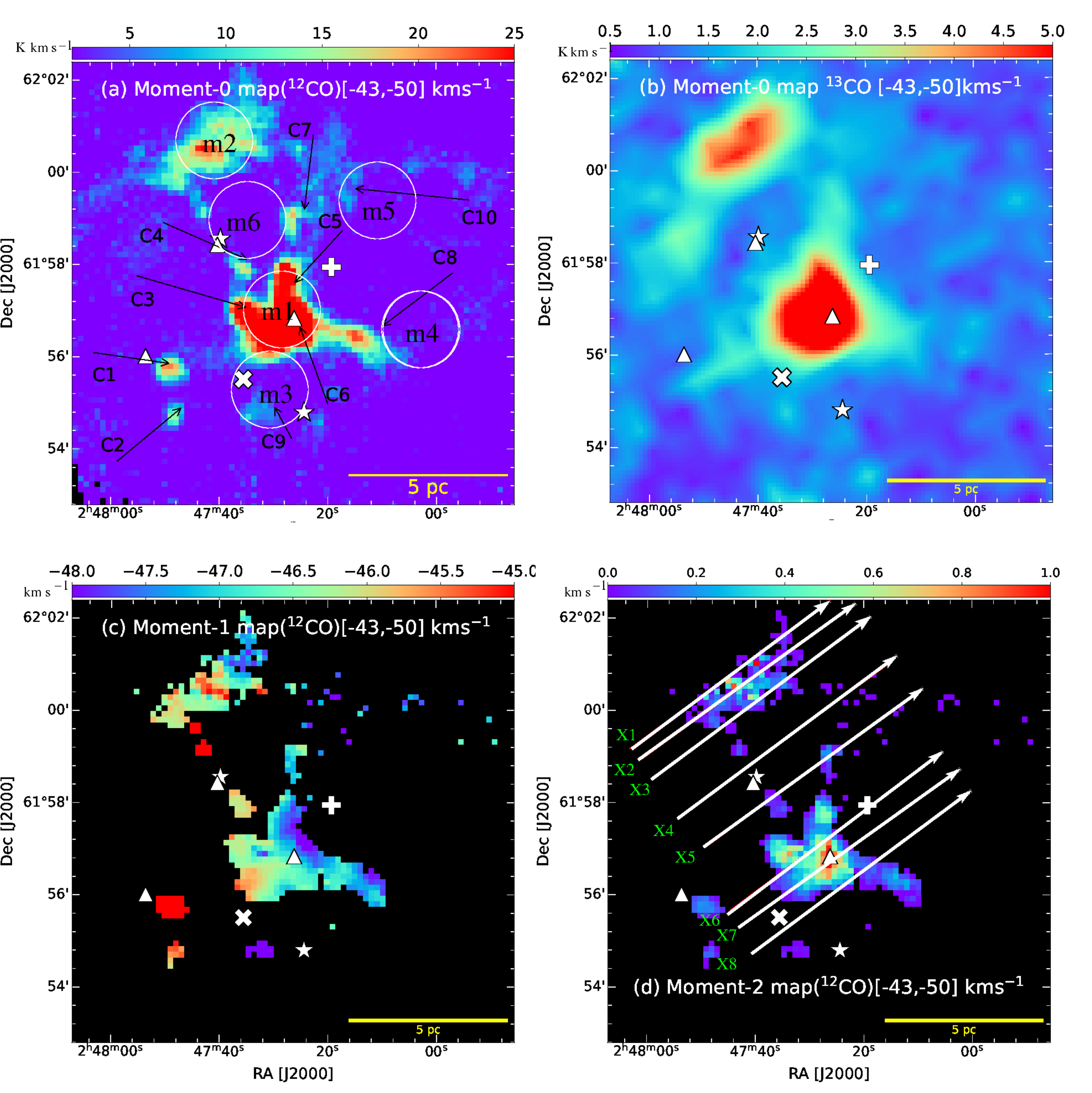}
\caption{\label{momap} a) $JCMT$ ($^{12}$CO(J =3$-$2)) moment-0 map along with the four regions (m1 to m4) from where we have extracted the spectra are shown with the red circles. Molecular clumps (C1 to C10) identified by \citet{2011AJ....141..123A} are also marked in the figure. (b) MWISP ($^{12}$CO(J =3$-$2)) moment-0 map. c) $JCMT$ ($^{12}$CO(J =3$-$2)) moment-1 map. d) $JCMT$ ($^{12}$CO(J =3$-$2)) moment-2 map, the map is overplotted with the lines along which PV maps are obtained (see; Figure \ref{pvd}). 
}
\end{figure*}

\begin{figure*}
\centering
\includegraphics[width=1\textwidth]{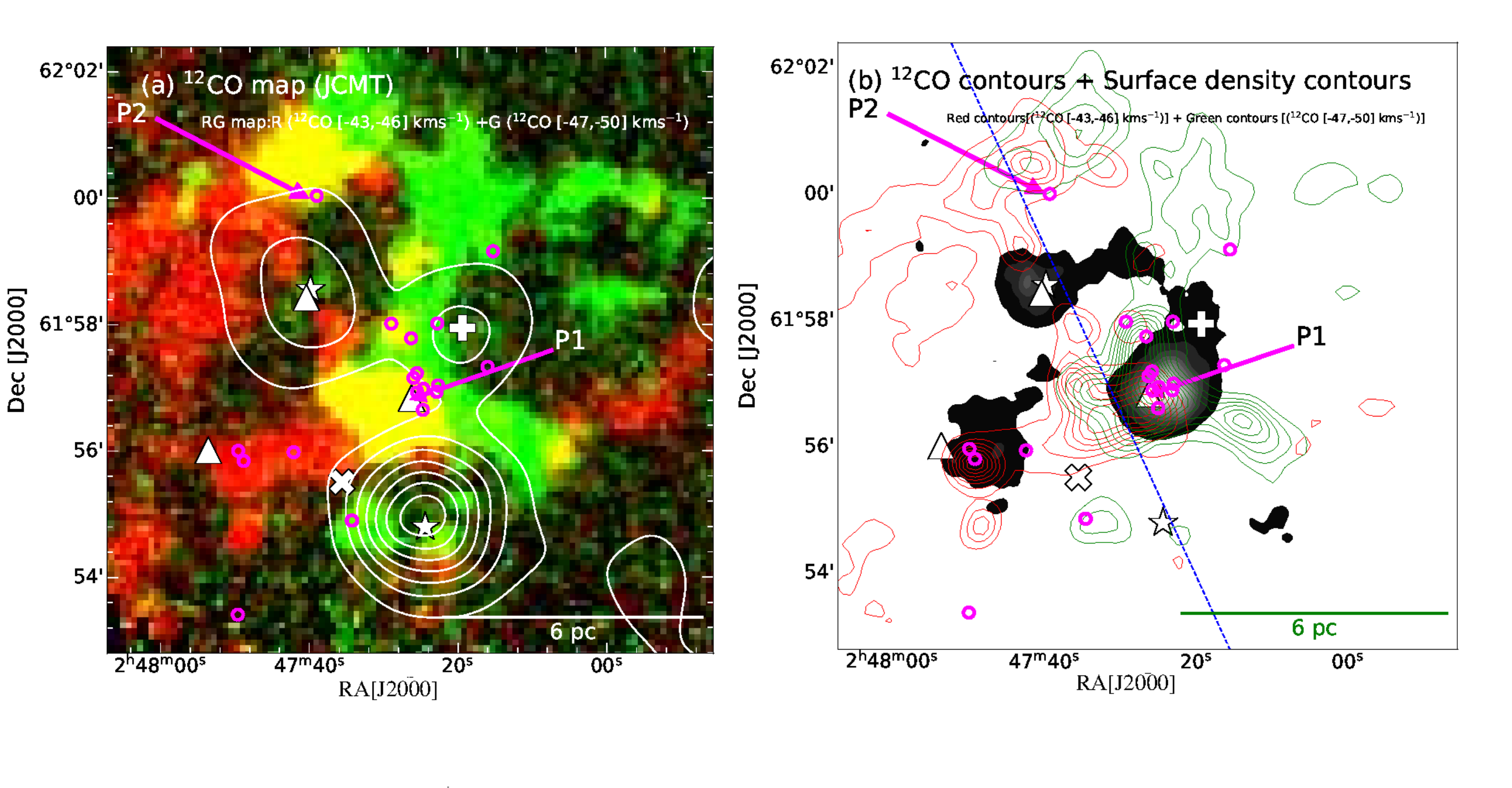}
\caption{\label{mom1} (a) Two-color composite image of S193 complex produced using $^{12}$CO(J =3$-$2) intensity map ((red:[-43,-46], green:[-47,-50] kms$^{-1}$), 
the $CGPS$ 1420 MHz contours are also shown with white color. (b) Filled contours showing the surface density map of the S193 complex, the map is also overplotted with the $^{12}$CO(J =3$-$2 )
contours showing two velocity components. The blue dashed line shows the possible axis of collision.
}
\end{figure*}


\subsubsection{Kinematics of the molecular gas in the S193 complex} \label{jcmK}

To trace the distribution of the molecular gas and determine the gas kinematics in the S193 complex, we have used the $^{12}$CO(J =3$-$2) $JCMT$ and 
MWISP $^{13}$CO(J =1$-$0) line data (cf. Section \ref{sec2}). In Figure \ref{spec}, we have shown the $^{12}$CO and $^{13}$CO velocity profiles at six different locations in the direction of the S193 complex (m1-m6; see Figure \ref{momap}(a)). In some of the locations we got a double-peaked velocity profile in the $^{12}$CO, these positions also host some of the molecular clumps identified by \cite{2011AJ....141..123A} . However, at positions P1 and P2 (where the intensity is maximum) we did not get a double-peaked velocity profile. In the positions (m1 and m2; cf. Figure \ref{momap}(a)) the shape of the $^{13}$CO profile almost mimics the $^{12}$CO while at positions m3, m4 we could see a double-peaked velocity profile of $^{13}$CO.
Overall, the velocity profiles indicate a velocity range of [-43,-50] kms$^{-1}$  with double peaks, i.e., at approximately -45 kms$^{-1}$ and -48 kms$^{-1}$ at few particular locations. These double-peaked velocity profiles could show two molecular clouds having different velocity components which might be interacting with each other \citep[see for example; ][]{2019ApJ...875..138D}.

In Figure \ref{momap} we have shown moment-0 (a), moment-1 (c), and moment-2 (d) map of $^{12}$CO(J =3$-$2). 
The intensity map (moment-0) is integrated over a velocity range of [-43,-50] kms$^{-1}$. 
The molecular gas shows fragmented morphology around the H\,{\sc ii} regions with cavities created by the feedback 
from the massive stars. We can see two prominent peaks in the intensity map, one at P1 and another at P2, where we have also seen peaks in the  $Herschel$ column density map (see Section \ref{clump}). The location of massive stars, IRAS sources, and the YSOs identified in the present study are also shown in the figure. Most of the YSOs and IRAS sources are distributed towards peak P1.
The ten molecular clumps identified by \citet{2011AJ....141..123A} are also shown in the figure and are well correlated with our moment-0 map. 

\citet{2011AJ....141..123A} have reported the masses and radii of the detected clumps to be in between 61-7 M$_\odot$ and 0.5-0.2 pc, respectively. These values are a bit lower than those  we have estimated using the Hershel column density maps. They have adopted a different value of the distance for the S193 complex, i.e., 2.96 kpc, than the present study (4.84 kpc). The most massive clump C3 (61$\pm$19 M$_\odot$), together with clumps C5 and C6, lie within the peak P1, whereas the least massive clump C2 (7$\pm$2 M$_\odot$), along with clump C1 is located towards the identified SEC clustering (cf. Figure \ref{momap}(a)). There appears to be some correlation between the dust and gas clumps as the most massive dust clumps are distributed towards peak P1 while the less massive clumps are scattered towards SEC (cf. Section \ref{clump}). In Figure \ref{momap}b, we have shown the first-moment (moment-1) map of $^{12}$CO (J =3$-$2), 
which is the intensity-weighted mean velocity of the emitting gas. 
The first-moment map shows two velocity components, i.e., one mainly in the eastern direction at [-43,-46] kms$^{-1}$ and another mainly in the western 
direction at [-47,-50] kms$^{-1}$, entangling at P1 and P2. The moment-2 map, which shows velocity dispersion, indicates relatively high values at P1 and P2.

Figure  \ref{momap}(b) shows the moment-0 map of $^{13}$CO(J =1$-$0). The $^{13}$CO gas is relatively optically thin and traces dense gas compared to the $^{12}$CO. The moment-0 map, integrated in a velocity interval of [-43,-50] kms$^{-1}$ peaks at P1 and P2 and shows almost the same morphology as $^{12}$CO despite having low resolution (cf. Section \ref{archiv}).

In Figure \ref{mom1}a, we show the color composite image of S193 complex generated from the $^{12}$CO maps at [-43,-46] and [-47,-50] kms$^{-1}$ in red and green colors, respectively. 
The molecular cloud near the massive star seems to be blown away from the massive stars, as indicated by the $CGPS$ radio contours located in the region of low brightness. 
In Figure \ref{mom1}b, we have shown the spatial connection of the two components, which is superimposed by the surface density map of the stars in the S193 complex (see Section \ref{group}). The stellar clustering, i.e., [BDS20003]57, and almost all the YSOs are located in the interaction zone P1.
 
To check whether there is a connection between these two components in velocity space, we have also obtained the position-velocity (PV) map of $^{12}$CO and $^{13}$CO along multiple lines. These lines are marked over the moment-2 map of $^{12}$CO (Figure \ref{momap}c) and $^{13}$CO (Figure \ref{mom1}d), most of these lines are passing through interaction zone P1 and P2. We obtained the PV map by choosing the width of the slices as 1$'$, which are shown in Figure \ref{pvd}. Most of the PV maps obtained using $^{12}$CO (Figure \ref{pvd}a) $^{13}$CO (Figure \ref{pvd}b) show a clear signature of velocity gradient where velocity is increasing while going from east to west direction. No apparent signature suggests that two groupings in velocity space are separated with a diffuse emission.

\begin{figure*}
\centering
\includegraphics[width=0.8\textwidth]{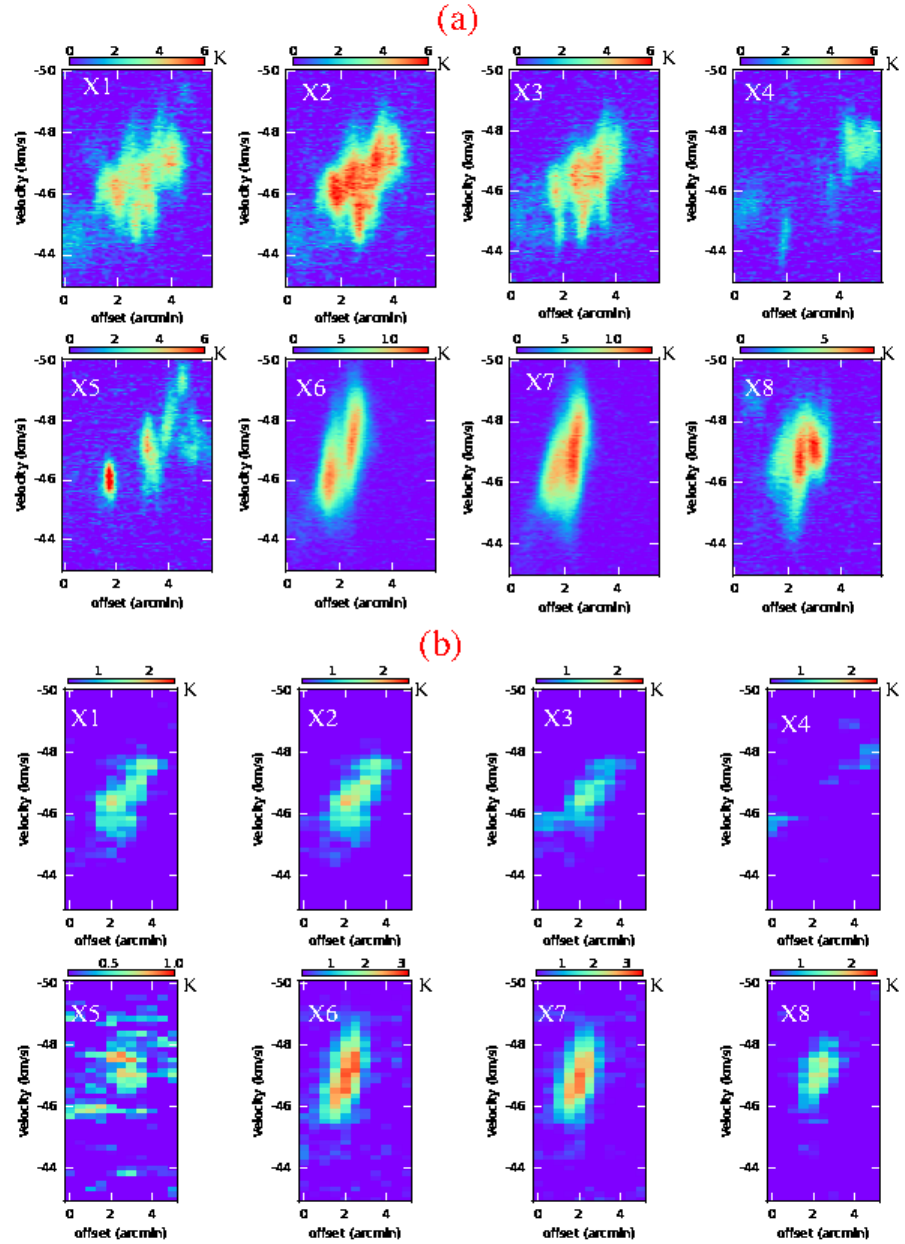}
\caption{\label{pvd} PV maps obtained using the $^{12}$CO(J =3$-$2) (a), and $^{13}$CO(J =1$-$0) (b) data, these maps are obtained along the different lines (white lines in Figure \ref{momap}d) as discussed in Section \ref{jcmK}.}
\end{figure*}


\subsubsection{Feedback of massive stars in the S193 complex} \label{sec:feedb}

We also examined the effect of feedback from the massive stars in the S193 complex by calculating total feedback pressure. Total feedback pressure consists of the stellar wind pressure (P$_{wind}$), radiation pressure (P$_{rad}$), and the H\,{\sc ii} region pressure (P$_{H\,{\sc ii}}$) \citep{2012ApJ...758L..28B,2017ApJ...851..140D}, which are given by the following equations \citep[see for details,][]{2012ApJ...758L..28B}.


\begin{equation}
P_{rad} = L_{bol}/ 4\pi c D_{s}^2; 
\end{equation}

\begin{equation}
P_{H\,{\sc ii}} = \mu m_{H} c_{s}^2\, \left(\sqrt{3N_{UV}\over 4\pi\,\alpha_{B}\, D_{s}^3}\right);   
\end{equation}

\begin{equation}
P_{wind} = \dot{M}_{w} V_{w} / 4 \pi D_{s}^2; 
\end{equation}

In the equations mentioned above, L$_{bol}$ denotes the bolometric luminosity of the ionizing source, $D_{s}$ is the distance at which pressure has to be calculated,  m$_{H}$ is the hydrogen atom mass,
c$_{s}$ is the sound speed in the photoionised region \citep[=11 km s$^{-1}$;][]{2009A&A...497..649B}. N$_{UV}$  in the above equation denotes the Lyman continuum flux, `$\alpha_{B}$' is the radiative recombination coefficient, and $\dot{M}_{w}$,  V$_{w}$ is the mass-loss rate, and the wind velocity of the ionizing source, respectively. 

Since there is ambiguity regarding the spectral type of the ionizing sources in the S193 complex, we are considering both the lower and the upper limits of the spectral type for pressure calculation. 
As a lower limit, we consider B2.5V as the spectral type of all the sources. B2.5V is the reported spectral type for the S193-1 and S192-1; for S194-1, it can also be treated as the lower limit.
We are adopting $L_{bol}$= 5012 L$_{\odot}$ \citep[][]{1973AJ.....78..929P}, $\dot{M}_{w}$ $\approx$ 10$^{-10}$ M$_{\odot}$ yr$^{-1}$, and V$_{w}$ $\approx$ 700 km s$^{-1}$ \citep[]{2019AJ....158...73K}, and N$_{UV}$ = 7.763 $\times$ 10$^{44}$ \citep[]{1973AJ.....78..929P} for the B2.5V spectral type. 
We obtained the total pressure ($P$= $P_{H\,{\sc ii}}$+$P_{rad}$+ $P_{wind}$) for the S192, S193, and S194 H\,{\sc ii} regions 
as 1.91 $\times$ 10$^{-12}$ dynes\, cm$^{-2}$, 1.37 $\times$ 10$^{-12}$ dynes\, cm$^{-2}$, and 4.08 $\times$ 10$^{-12}$ dynes\, cm$^{-2}$, respectively. If we add these values to get the pressure due to all three H\,{\sc ii} regions at the position of the central cluster [BDS2003]57 ($P_{all}$= $P_{192}$+$P_{193}$+$P_{194}$), it comes out to be $\sim$ 7.36 $\times$ 10$^{-12}$ dynes\, cm$^{-2}$.
For the upper limit, we consider the spectral type as B0.5, which is suggested by our radio calculation (cf. Section \ref{ioni}).
For a B0.5 star, we adopted the values of $L_{bol}$= 19952 L$_{\odot}$, $\dot{M}_{w}$ $\approx$ 2.5$\times$ 10$^{-9}$ M$_{\odot}$ yr$^{-1}$,  V$_{w}$ $\approx$ 1000 km s$^{-1}$ \citep[]{2017ApJ...834...22D}, and N$_{UV}$ = 7.763 $\times$ 10$^{44}$ \citep[]{1973AJ.....78..929P}. We obtained the total pressure ($P$= $P_{H\,{\sc ii}}$+$P_{rad}$+ $P_{wind}$) for the S192, S193 and S194 H\,{\sc ii} regions 
as 2.78 $\times$ 10$^{-11}$ dynes\, cm$^{-2}$, 2.03 $\times$ 10$^{-11}$ dynes\, cm$^{-2}$, and 5.6 $\times$ 10$^{-11}$ dynes\, cm$^{-2}$, respectively. The pressure due to all three H II regions comes out to be $\sim$ 1.06 $\times$ 10$^{-10}$ dynes\, cm$^{-2}$. Here it is worthwhile to note that the distance ($D_{s}$) is estimated without taking into consideration of the projection effects. Thus, the estimated distance represents its lower limit and it will lead to the upper limit in the estimated pressure values.
\section{Discussion} \label{sec4}
The S193 complex has complex and diverse morphological features like young star clusters, three H\,{\sc ii} regions, a few massive stars, and several molecular clumps. Figure \ref{fig1} shows the distribution of the massive stars, dust and gas, PDRs, and ionised gas along with the YSOs in the S193 complex. The presence of YSOs indicates the ongoing recent star formation activity in the region. In this section, we have discussed the implications of the observed morphological features in the S193 complex and investigated the star formation scenario.

Most YSOs in the S193 complex are distributed toward the central region populating the young star cluster [BDS2003]57.
The other young star cluster T162 is associated with the H\,{\sc ii} region S193 and contains a massive star (S193-1). 
The new clustering identified in the south-west direction (`SEC') has few YSOs. We have estimated the MF slope for the central cluster [BDS2003]57 from our deep NIR data
and found a change of MF slope from the high to low mass end with a turn-off at around 1.5 M$_\odot$. This truncation of MF slope at a bit higher mass bins has often been noticed in other \textbf{star-forming regions (SFRs)} under the influence of massive OB-type stars \citep{2020ApJ...891...81P,2008MNRAS.383.1241P,2017MNRAS.467.2943S, 2007MNRAS.380.1141S, 2008MNRAS.384.1675J}.
The higher-mass stars mostly follow the Salpeter MF value, i.e., $\Gamma$=-1.35  \citep{1955ApJ...121..161S}.
At lower masses, the IMF is less constrained but appears to flatten
below 1 M$_\odot$  and exhibits fewer stars of the lowest masses \citep{2016ApJ...827...52L,2015arXiv151101118L,2003PASP..115..763C,2002Sci...295...82K}.
While the higher-mass domain is thought to be mostly formed through fragmentation and/or accretion onto
the protostellar core \citep[e.g.,][]{2006MNRAS.370..488B,2002ApJ...576..870P}
in the low-mass and substellar regime, additional
physics is likely to play an important role. The density, velocity
fields, chemical composition, tidal forces in the natal molecular
clouds, and photo-erosion in the radiation field of massive stars in
the vicinity can lead to different star formation processes and
consequently, some variation in the characteristic mass (turn-off point) of the IMF \citep{2009MNRAS.392.1363B,2005MNRAS.356.1201B,2004A&A...427..299W,2002ApJ...576..870P}.

\begin{figure*}
\centering
\includegraphics[width=0.7\textwidth]{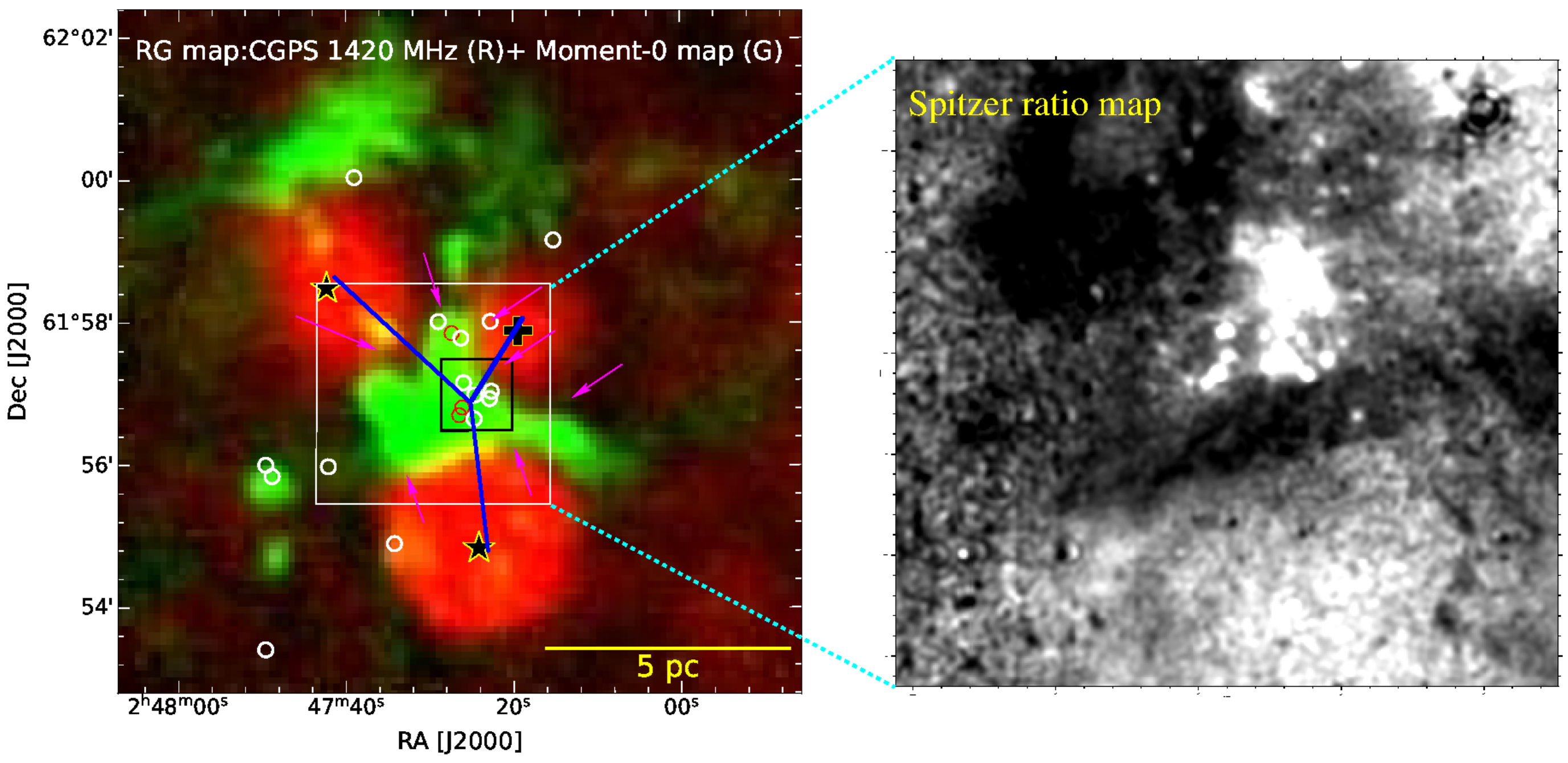}
\caption{\label{feeb} Two colour-composite image produced using CGPS 1420 MHz radio continuum image (red) and  $^{12}$CO(J =3$-$2) moment-0 map (green). In this figure, massive stars S193-1 and S193-2 are shown with the star symbol, while candidate massive star S194-1 is shown with the plus symbol. The black box represents the cluster's core region [BDS2003]57, white circle are the YSOs, and blue lines connect the cluster with the massive stars. The $Spitzer$ ratio map (see Figure \ref{fig1}c) enclosed within a white box is shown in the right panel.}
\end{figure*}

Each cluster identified in the S193 complex has an associated IRAS source. We saw the presence of cold gas and dust clumps at the location 
of [BDS2003]57 and SEC in the $Herschel$ and molecular maps. 
The [BDS2003]57 cluster is associated with the peak in the $Herschel$ 500 $\mu$m emission and traced with a relatively \textbf{low} 
temperature ($\sim$14 K) in the $Herschel$ temperature map. We have also seen molecular line emission
(a bright region in the IRAC ratio map) from the [BDS2003]57. By analysing the $^{12}$CO(J =3$-$2) data, we have identified the presence of two velocity components in the S193 complex at
velocities around -45 kms$^{-1}$ and -48 kms$^{-1}$.
 Two spatially overlapping zones of clouds, P1 and P2 (see Figure \ref{momap}a), are observed, which are peaking at the intensity maps of the $^{12}$CO(J =3$-$2), $^{13}$CO(J =1$-$0) emission. Zone P1 has a young open cluster [BDS2003]57, YSOs, a cold gas clump, a massive molecular clump (C1).
We also saw three more molecular clumps (C3, C5, and C6) lying inside this overlapping zone P1. Zone P2 also
have two massive $Herschel$ clumps (C2, C7; see Section \ref{clump}) along with a YSO.
Though, in the $^{12}$CO and $^{13}$CO intensity map (see Figure \ref{momap}a, \ref{momap}b), the space near the massive stars seems to be almost devoid of gas but are 
in very close proximity of overlapping zones P1 and P2. The massive stars evolve faster than the low mass stars, affecting their natal
environment very soon. In that way, the space near the massive stars does not always give a picture of the initial physical conditions. We show a blue dashed 
line in Figure \ref{mom1}b, which we can consider as the axis of two clouds colliding.  
This axis passes through the overlapping zones P1 and P2 and
covers the positions of massive stars in the S193 complex. Few cavities are visible along this axis around massive stars (cf. Figure \ref{mom1}b), filled with ionised gas and surrounded by the PDRs, dust, and gas. These findings lead us to explore the CCC process in the S193 complex, which strongly favours the massive dense core, young clusters, and YSOs formation at the overlapped region of the two interacting clouds \citep{2017ApJ...835..142T, 2018PASJ...70S..44F,2019ApJ...875..138D,2021PASJ...73S.405F}.
Many authors list the observational signatures of the CCC \citep{2017ApJ...835..142T,2019ApJ...875..138D}. An important signature of CCC is
the existence of a compressed layer of gas at the junction of two clouds, known as the bridge-like feature \citep{2015MNRAS.450...10H, 2015MNRAS.454.1634H, 2017ApJ...835..142T, 2019ApJ...875..138D}.
The bridge feature shows the connection of two clouds in velocity space, which is observationally seen as groupings in the PV maps separated by diffuse emission \citep{2015MNRAS.450...10H,2015MNRAS.454.1634H}.
We have shown PV maps along different lines in our $^{12}$CO and $^{13}$CO maps to investigate the bridge feature. The $^{12}$CO(J =3$-$2) transition is known to be optically thin as compared to other $^{12}$CO
transitions such as J=1-0 and J=2-1. It is a very useful tracer of medium density (10$^{4}$ at 20K) material, while $^{13}$CO is a tracer of the dense gas. In all of the PV maps (except x1 and x2; see Figure \ref{pvd})}
for $^{12}$CO (a) and $^{13}$CO (b),  one can see the velocity gradient from east to west but cannot see any significant feature suggesting the velocity connection of two clouds. Another important observational signature of the
CCC is the spatial fit of enhanced intensity and depressed region \citep{2015MNRAS.454.1634H, 2022ApJ...934....2M}. Intensity enhancement is termed `key' while intensity depressed feature is known as `key-hole.'
In Figure \ref{mom1}a, where we have plotted two velocity components with red and green colours, this kind of distribution is not observable. However, the spatial connection of two velocity components with massive stars and clustering 
found at the interface suggests the CCC in this region; other signatures of CCC are not observed with the existing data sets. The possibility of not seeing the observational signature of CCC could be that feedback from massive stars, might be responsible for diminishing these important features as three evolved H\,{\sc ii} regions are distributed towards the S193 complex \citep{2022ApJ...934....2M}. Another possibility suggesting the influence of feedback from the massive \textbf{stars} 
played a role in forming the central cluster [BDS2003]57 is discussed in the following paragraph.\\

The S193 complex contains a few massive stars, and feedback from them (intense UV radiation, stellar wind) could also be the reason 
behind the observed velocity gradient in the S193 complex. The massive stars in the S193 complex strongly affect their surroundings, evident with the PDR structures and the cavities 
formed in the cloud. Interestingly the central cluster [BDS2003]57 is surrounded by the H\,{\sc ii} regions S192 (from the south), S193 (from the north-east), and S194 (from the north-west), as we can see
radio emissions from the H\,{\sc ii} regions surrounding the cluster (see Figure \ref{feeb}). To investigate the effect of feedback from the massive stars in the cluster, we have shown two colour composite 
image of the S193 complex made using CGPS 1.42 GHz radio continuum image (red) and the $^{12}$CO(J =3$-$2) moment-0 map (green) in Figure \ref{feeb}. Let us take a close look at the central region hosting the 
cluster. It is visible that this region shows curved morphology from all three sides where radio continuum emission interacts with the molecular cloud (shown with magenta arrows in Figure \ref{feeb}).
 This signature strongly suggests the compression of the cloud due to the interaction with the radio emission; interestingly, we have also seen PDR structures at the interaction zone, further strengthening this
 idea. This central region also hosts molecular clumps, dust condensations, YSOs, and the cluster. The compression of the molecular cloud from the sides of all three H\,{\sc ii} regions could be responsible 
 for the formation of clumps and, subsequently, the cluster [BDS2003]57, which we can attribute to the positive feedback of the massive stars.
To quantitatively examine this scenario, we also calculated the pressure of the massive star at the position of the cluster [BDS2003]57; the pressure calculations had been given as one of the main arguments supporting the 
feedback scenario \citep{2020ApJ...891...81P}. Considering the ambiguity regarding the spectral type of the massive stars in the region, we have calculated the upper and lower limit of the total pressure due to all three H\,{\sc ii}
regions at the position of [BDS2003]57. The lower limit of the total pressure (7.36 $\times$ 10$^{-12}$ dynes\, cm$^{-2}$) is similar to the value of a cool molecular cloud ($P_{MC}$$\sim$10$^{-11}$--10$^{-12}$ dynes cm$^{-2}$ for a 
temperature $\sim$20 K and particle density $\sim$10$^{3}$--10$^{4}$ cm$^{-3}$) \citep[see Table 7.3 of][]{1980pim..book.....D}. However, an upper limit of the total pressure (1.06 $\times$ 10$^{-10}$ dynes\, cm$^{-2}$) is greater than
the value of a typical cloud. If taken into consideration, the upper limit of the pressure suggests that the massive stars' feedback has triggered the cloud's collapse; however, a detailed spectroscopic analysis of the massive 
stars would shed greater light in this regard. 
In our analysis, we found that the dynamical age of the H\,{\sc ii} regions S192, S193, and S194, for ambient density n$_0$=10$^3$(10$^4$) cm$^{-3}$ are 0.6 (1.9) Myr, 0.6 (2.0) Myr, 0.4 (1.4) Myr, respectively. The dynamical age of these H\,{\sc ii} regions considering the ambient density of n$_0$=10$^4$ cm$^{-3}$ seems to be high enough to trigger the formation of the YSOs which are considered to have an average age of 0.46 (Class I) and 1-3 Myr (Class II) \citep{2009ApJS..181..321E}. Since we are proposing that the star formation in the central cluster [BDS2003]57 is happening due to the compression of existing material from the surrounding H\,{\sc ii} regions. In this case, considering the ambient density at the higher end is a good approximation and suits the triggered star formation scenario. 

\section{Summary and conclusions} \label{sec5}

In the paper, we aimed to understand the formation of massive stars and young star clusters in the S193 complex. We probed the physical environment around the S193 complex by carefully examining the multiwavelength infrared data and the molecular line data. 
We present observational findings and conclusions drawn from the above studies as follows. 

\begin{itemize}

\item
We have identified three clustering of stars in the S193 complex. The clustering contains previously identified young open clusters and a new grouping of stars in the south-west direction. The membership of the identified clustering is constrained using the Gaia-DR3 data, and the
clusters seem connected in the proper-motion space. The distance of the S193 complex is estimated as 4.84 kpc from the PM data.

\item

The distribution of dust in the S193 complex is traced using the MIR and FIR images. The dust is distributed in arc-type structures around the H\,{\sc ii} regions. The PDRs are also traced around the H\,{\sc ii} regions, indicating the significant heating/impact of the massive stars. We also traced Br$\alpha$ emission from the H\,{\sc ii} regions, whereas a prominent hydrogen line emission towards the cluster [BDS2003]57 is observed.

\item
Using the observed NIR data and the MIR data from the $Spitzer$, we identified 27 YSOs in the S193 complex. Out of the total YSOs, two 
are Class\,{\sc i}, and twenty five are Class\,{\sc ii} objects. Most of the YSOs are distributed towards the cluster [BDS2003]57. We have also identified 16
molecular clumps in the S193 complex using the $Herschel$ column density map. The most massive clump is seen towards the central cluster, [BDS2003]57, accompanying 
several YSOs. This clump is relatively cold and traced with a temperature $\sim$14 K in the $Herschel$ temperature map.

\item
Using our observed NIR data, we have also calculated the slope of the MF for the central cluster [BDS2003]57. We found a break in the slope of the MF at 1.5 M$_\odot$. In the mass range of $3>M_\odot>1.5$ it is found to be $-2.46\pm0.01$ while in the mass range of $1.5>M_\odot>0.3$, it is found as $+0.64\pm0.20$.

\item
The distribution of the ionised gas in the S193 complex is examined using the H$\alpha$ and the $CGPS$ 1420 MHz images. Using radio flux, we further constrained the spectral type (B0.5-B1) of the ionizing sources for H\,{\sc ii} regions S192, S193, and S194. 

\item
We traced two velocity components towards the S193 complex with velocities [-43,-46] and [-47,-50] km/s. Two prominent spatially overlapping zones, P1 and P2
peaking in the intensity map were traced towards the center and right-above the H\,{\sc ii} region S193 complex. The overlapping zone P1 accompanies the [BDS2003]57 clustering, cold gas clump, YSOs.

\item
We explored the CCC scenario in the S193 complex and it appears that the cause of the formation of a new generation of stars in this region is not the CCC process but the feedback from the massive stars.

\item
We have investigated feedback's possible role in forming the central cluster [BDS2003]57 by calculating the total pressure and inspecting the physical environment around the central clump. It suggests that the massive stars could have played a role in forming the central cluster, although a detailed spectroscopic and age analysis is needed to shed more light on this aspect.   

\end{itemize}

\section*{Acknowledgments} 
\textbf{We thank the anonymous referee for the constructive and valuable comments that helped to improve the manuscript.} We thank the staff at the 3.6m DOT, Devasthal (ARIES) and IR astronomy group at TIFR, for their cooperation during TANSPEC observations. We also acknowledge the TIFR Near Infrared Spectrometer and Imager mounted on 2 m HCT using which we have made NIR observations. This research made use of the data from the Milky Way Imaging Scroll Paint-
ing (MWISP) project, which is a multi-line survey in 12CO/13CO/C18O along the northern galactic plane with PMO$-$13.7m telescope. We are grateful to all the members of the MWISP working group, particularly the staff members at PMO-13.7m telescope, for their long-term support. MWISP was sponsored by National Key R\&D Program of China with grant
2017YFA0402701 and by CAS Key Research Program of Frontier Sciences with grant QYZDJ-SSW-SLH047. D.K.O. acknowledges the support of the Department of Atomic Energy, Government of India, under Project Identification No. RTI 4002.

\section*{Data availability}
The data underlying this article will be shared on reasonable request to the corresponding author.


\bibliography{Rakesh}{}
\bibliographystyle{mnras}

\clearpage

\begin{table*}
\centering
\caption{\label{stats} Statistics of the NIR observations}
\begin{tabular}{@{}rcccc@{}}
\hline
Band& Number of & Detection &  \\
    &   sources & Limit (mag) \\
\hline
$J$    &1341$^b$+ 168$^c$      &18.70$^b$+20.347$^c$ \\      
$H$    &1838$^b$+ 204$^c$      &18.44$^b$+20.149$^c$ \\
$K$    &1721$^b$+ 185$^c$      &17.79$^b$+18.319$^c$  \\
\hline
\end{tabular}

$^b$: data from $TIRSPEC$ ($10^\prime\times10^\prime$ FOV);\\
$^c$: data from $TANSPEC$ ($1^\prime\times1^\prime$ FOV);\\
\end{table*}

\begin{table*}
\centering
\caption{\label{archilog} Details of the archival data set used in the present study.}
\footnotesize
\begin{tabular}{@{}lccccc@{}}
\hline
\hline
Survey & Details \\
      &       &  \\
\hline
The INT Photometric H$\alpha$ Survey of the Northern Galactic Plane (IPHAS) & \citep[H$\alpha$; resolution $\sim$1$''$;][]{2005MNRAS.362..753D} \\
The Pan-STARRS1 Surveys (PS1)         & \citep[0.4-1.2 $\mu$m;][]{2016arXiv161205560C} \\
Canadian Galactic Plane Survey ($CGPS$) & \citep[1.42 GHz; resolution $\sim$1$'$$\times$1$'$csc$\delta$;][]{2003AJ....125.3145T} \\
NRAO VLA Sky Survey  & \citep[NVSS; $\lambda$ =21 cm; resolution $\sim$46$''$;][]{1998AJ....115.1693C} \\
Herschel infrared Galactic Plane Survey (Hi-Gal) & \citep[$\lambda$ =70 - 500 $\mu$m; resolution$\sim$5.8 -46$''$;][]{2010PASP..122..314M}\\ 
Warm-{\it Spitzer} GLIMPSE360 Survey    & \citep[$\lambda$ =3.6 -4.5 $\mu$m; resolution $\sim$2$''$;][]{2005ApJ...630L.149B} \\
$JCMT$ Science Archive ($^{12}$CO(J =3$-$2)) & \citep[345GHz; resolution $\sim$14$''$; PROJECT = `M08BH15';][]{2009MNRAS.399.1026B}\\
Milky Way Imaging Scroll Painting  ($^{13}$CO(J =1$-$0)) & \citep[MWISP, resolution $\sim$55$''$;][]{2019ApJS..240....9S} \\
Two Micron All Sky Survey  & \citep[2MASS, $\lambda$ =1.1 - 2.2 $\mu$m; resolution $\sim$2.5$''$;][]{2006AJ....131.1163S} \\
Wide-field Infrared Survey Explorer & \citep[WISE, $\lambda$ =3.6 - 22 $\mu$m; resolution $\sim$6 -12$''$;][] {2010AJ....140.1868W}  \\
\hline
\end{tabular}
\end{table*}

\begin{table*}
\centering
\label{pmt}
\caption{\label{PMT} Sample of 80 stars identified as a member of the S193 complex. 
The complete table is available in the electronic form only.}
\begin{tabular}{ccrcccccc}
\hline
ID& $\alpha_{(2000)}$&$\delta_{(2000)}$& Parallax$\pm\sigma~~~~~$&$\mu_\alpha\pm\sigma$&$\mu_\delta\pm\sigma$ & $G$ & $G_{BP}-G_{RP}$ & Probability\\
& {\rm $(degrees)$} & {\rm $(degrees) $} & (mas)&  (mas/yr)& (mas/yr) & (mag) & (mag) & (Percentage)\\
\hline
1 & 41.885990 & 61.909958 &  $0.065\pm 0.037$ &$-0.345\pm 0.016$ &  $0.139\pm 0.035$ &  15.842  &  1.262  &  100 \\
2 & 41.916260 & 61.923225 &  $0.262\pm 0.257$ &$-0.397\pm 0.112$ &  $0.153\pm 0.225$ &  19.020  &  1.616  &   98 \\
3 & 41.906441 & 61.927036 &  $0.444\pm 0.340$ &$-0.198\pm 0.137$ &  $0.416\pm 0.314$ &  19.438  &  2.344  &   93  \\
4 & 41.916271 & 61.928421 &  $0.211\pm 0.133$ &$-0.328\pm 0.055$ &  $0.034\pm 0.117$ &  18.057  &  1.333  &   99   \\
\hline
\end{tabular}
\end{table*}

\begin{table*}
\centering
\caption{\label{cd1} Clumps identified in the S193 complex using the $Herschel$ column density map.
Center coordinates, radius (column 3) and mass (column 5) corresponding to each clump are tabulated below.}
\begin{tabular}{@{}lcccrc@{}}
\hline
ID & $\alpha_{(2000)}$&$\delta_{(2000)}$&Radius& Mass \\
 & {\rm $(^h:^m:^s)$} & {\rm $(^o:^\prime:^{\prime\prime)} $} &  (pc) & (M$_\odot$) \\
\hline
  1   &  02:47:27.1 & 61:56:43  &	  1.5   &   1141.9 \\
  2   &  02:47:41.5 & 62:00:13  &	  1.8   &   912.2  \\
  3   &  02:47:49.2 & 61:55:43  &	  1.2   &   342.7  \\
  4   &  02:47:16.0 & 61:56:19  &	  1.3   &   444.6  \\
  5   &  02:47:27.9 & 61:58:43  &	  0.9   &   227.5  \\
  6   &  02:47:36.4 & 61:56:49  &	  1.0   &   268.2  \\
  7   &  02:47:55.2 & 62:00:19  &	  1.5   &   527.8  \\
  8   &  02:47:35.6 & 61:57:49  &	  0.9   &   200.9  \\
  9   &  02:47:49.1 & 61:54:37  &	  1.3   &   334.0  \\
  10  &  02:47:30.5 & 61:59:31  &	  1.2   &   332.1  \\
  11  &  02:47:35.5 & 61:53:01  &	  1.5   &   431.5  \\
  12  &  02:47:16.8 & 61:58:55  &	  0.9   &   174.1  \\
  13  &  02:47:33.0 & 61:54:43  &	  1.0   &   200.5  \\
  14  &  02:47:16.8 & 61:59:43  &	  0.9   &   193.7  \\
  15  &  02:47:24.5 & 61:54:31  &	  0.9   &   161.7  \\
  16  &  02:47:20.3 & 61:55:01  &	  1.1   &   224.6  \\
\hline
\end{tabular}
\end{table*}

\clearpage

\appendix

\bsp    
\label{lastpage}
\end{document}